\documentclass[5p]{elsarticle} % seleccionar: preprint, review, 1p, 3p, 5p

\usepackage{lineno}
\modulolinenumbers[5]
\usepackage{mathtools}
\journal{ }

\usepackage{placeins}
\usepackage{url}
\usepackage[colorlinks=true, citecolor=blue, linkcolor=blue, filecolor=blue,urlcolor=blue]{hyperref}

\begin{document}

\begin{frontmatter}

\title{Using validated reanalysis data to investigate the impact of the PV system configurations at high penetration levels in European countries}

\author[mymainaddress]{Marta Victoria\corref{mycorrespondingauthor}}
\ead{mvp@eng.au.dk}
\author[mymainaddress]{Gorm B. Andresen}

\cortext[mycorrespondingauthor]{Corresponding author}

\address[mymainaddress]{Department of Engineering, Aarhus University, Inge Lehmanns Gade 10, 8000 Aarhus C, Denmark}

\begin{abstract}
Long-term hourly time series representing the PV generation in European countries have been obtained and made available under open license. For every country, four different PV configurations, \textit{i.e.} rooftop, optimum tilt, tracking, and delta, have been investigated. These are shown to have a strong influence in the hourly difference between electricity demand and PV generation. To obtain PV time series, irradiance from CFSR reanalysis dataset is converted into electricity generation and aggregated at country level. Prior to conversion, reanalysis irradiance is bias corrected using satellite-based SARAH dataset and a globally-applicable methodology. Moreover, a novel procedure is proposed to infer the orientation and inclination angles representative for PV panels based on the historical PV output throughout the days around summer and winter solstices. A key strength of the methodology is that it doesn't rely on historical PV output data. Consequently, it can be applied in places with no existing knowledge of PV performance.
\end{abstract}

\begin{keyword}
energy modeling, reanalysis, time series, clearness index, duck curve
%\texttt{elsarticle.cls}\sep \LaTeX\sep Elsevier \sep template
%\MSC[2010] 00-01\sep  99-00
\end{keyword}

\end{frontmatter}

%\linenumbers

\section{Introduction}
\label{intro}

The cost decrease experienced by photovoltaic (PV) solar energy throughout the last decade has been so dramatic that the projected installed capacities have been persistently underestimated by almost every relevant actor, \textit{e.g.}, the International Energy Agency (IEA) \cite{Fell_2015} or Greenpeace \cite{Creutzig_2017}. In 2016 and 2017, PV was the technology with the highest installed capacity among the renewable energy sources, and its cumulative installed capacity world-wide reached 390 GW at the end of 2017 \cite{IRENA_2018}.

\

With very low capital and operational costs and using a resource widely available, PV is seen as one of the main key enabling technologies in the transition towards a low-carbon energy system in almost every country. In fact, energy modeling efforts aiming to attain low cost highly renewable penetration in Europe include a significant share of PV generation. This result is obtained both when modeling the power system \cite{Eriksen_2017, Schlachtberger_2017} or when taking into consideration its coupling with other sectors \cite{Brown_2018, Connolly_2016}. 
PV also plays a major role when individual countries are modeled. Since the literature is vast in this case, the reader is referred to two papers in which a large number of countries are investigated with a consistent approach \cite{Breyer_2017, Jacobson_100_2017}. 

\

Since most of the energy models use PV generation time series as inputs and PV is expected to play a prominent role, we must ensure that its representation is as accurate as possible to reduce modeling uncertainties. In fact, the generation, bias correction and validation of PV time series at national scale lay within one of the main challenges for PV research in the near future identified by Kurtz \textit{et al.} \cite{Kurtz_2017}, namely ``research enabling more effective integration of solar electricity into the grid at high penetration levels''.
Moreover, Schlachtberger \textit{et al.} \cite{Schlachtberger_2018} showed that using different time series for renewable generation may significantly impact the energy model outcomes.

\

Hourly capacity factors are typically employed to deal with the inherent variability of renewables. For every hour, the capacity factor is calculated as the ratio between the delivered power and the cumulative installed capacity, \textit{i.e.}, the rated power of that technology. To obtain PV hourly capacity factors representative for a certain country two main strategies can be followed. First, historical data comprising generated electricity and installed capacity can be used to compute the hourly capacity factors. However, national Transmission System Operators (TSO) do not provide this information for every country. When they do, it would be, to some extent, based on  modeling since monitoring a myriad of small PV installations is in practice impossible. An additional drawback of this approach is, of course, that it does not allow to obtain hourly capacity factors for those countries where there is no capacity installed at the moment. The second strategy consists in using irradiance and temperature time series, together with a model for the PV systems, to compute hourly capacity factors. The resulting time series could be later bias-corrected using historical data when available. In this case, using accurate irradiance data is key to obtain proper PV time series. The necessary historical irradiance time series can be obtained from different sources: measurements at ground stations, satellite images or reanalysis (\textit{i.e.}, the output of global atmospheric simulations).

\

Using ground measurements entails some difficulties. First, the accuracy of the irradiance estimated on a certain location highly depends on the distance to the nearest ground station. Second, low quality or poor maintenance of measuring instruments, mainly pyranometers and pyrheliometers, could originate a significant bias in the estimated irradiance \cite{Gueymard_2009, Urraca_2017}. Hence, enabling the calculation of PV time series at country level using ground measurements requires a well-maintained and fine-meshed network of ground stations. This is not available for most countries.

\

The Joint Research Center (JRC) recently published the European Meteorological HIgh resolution RES time series (EMHIRES) dataset including PV hourly capacity factors for 30 years (1986-2015) at different aggregation levels: country, power market bidding zone, and territorial units NUTS1 and NUTS2 \cite{EMHIRES}.  EMHIRES uses the Meteosat-based SARAH (Surface Solar Radiation Database - Heliostat) \cite{SARAH} from CM-SAF (Satellite Application Facility on Climate Monitoring), and calculate irradiance on a tilted plane and convert it into PV generation using the PVGIS model \cite{PVGIS}. For every country, the modeled duration curve is corrected using a correction curve comprising 8760 values determined, for every hour, as the difference between the modeled and the TSO reported historical duration curve. This approach most probably leads to a significant overfitting. In general and prior to correction, the modeled EMHIRES time series overestimated the historical values, particularly for high capacity factors. This is most probably due to the fact that EMHIRES assumes the same configuration, inclination equal to 30$^\circ$ and south orientation, for every panel in a country.

%The goodness of the EMHIRES methodology is hard to assess since the authors, first, correct the modeled time series for every country in 2015 using an hourly calibration curve calculated using PV generation reported by TSOs and, later, use this same measured time series to evaluate the errors in the corrected time series.

\

Pfenninger and Staffell \cite{Pfenninger_2016} used the SARAH satellite dataset as well as MERRA-2 (Modern-Era Retrospective Analysis for Research and Applications) reanalysis dataset to obtain hourly capacity factors for European countries and make their results available through the very convenient Renewables Ninja website \cite{ninja}. To correct the bias of both modeled time series, Europe-wide multiplicative scaling factors are obtained by computing the average difference (modeled minus measured capacity factors) for all the individual sites whose data is available. The main limitation of this approach is the fact that the availability of historical data for individual sites is restricted to a few European countries and, consequently, the scaling factor applied to the whole Europe could be impact by the local climate of those countries. The authors also processed metadata for a large number of individual PV sites to estimate the configuration (orientation and tilt angle) of the installations. Moraes \textit{et al.} \cite{Moraes_2018} carried out a comparison on wind and PV time series from EMHIRES and Renewables Ninja in terms of correlation among hourly capacity factors, duration curves, annual full loads, weakly average, and seasonal ratios. Time series for 5 countries were compared for the period from year 2012 until 2014. Greater similarity between TSOs time series and EMHIRES was found for PV, however, the authors stress the significant differences among available time series and encourage future works validating and comparing them with actual renewable generation in additional years and countries. Lingfors \textit{et al.} developed PV time series for Sweden and validated them against historical data \cite{Lingfors_2016}. They used a local reanalysis model (STR{\AA}NG) to derive irradiance and discarded using the global reanalysis MERRA due to the lower accuracy attained. There are other previous works that use reanalysis datasets to estimate PV output time series but either they don't mention any validation procedure \cite{Haller_2012, Heide_2010} or the validation is limited to one country, Czech Republic in \cite{Jurus_2013} and Germany in \cite{Huber_2014}. 

\

When comparing to the historical PV output of individual sites, Pfenninger and Staffell \cite{Pfenninger_2016} found that modeled series using satellite-based or reanalysis dataset as input have a systematic error of the same order but the former showed lower RMSE \footnote{Root Mean Square Error (RMSE) calculate on the hourly capacity factors.}. Satellite-based datasets are known to capture better the local atmospheric phenomena and, in particular, the clouds dynamics which directly influence the irradiance at ground level \cite{Urraca_2017, Boilley_2015}.  In addition, SARAH dataset provides values for direct and global irradiance at ground level avoiding the modeling of diffuse irradiance that is required for reanalysis data and, hence, reducing uncertainties. Nevertheless, using irradiance from global reanalysis datasets to generate PV time series has significant advantages. Firstly, since reanalysis datasets cover the entire globe, the methodology can be replicated to obtain hourly capacity factors in every country no matter whether ground measurements or satellite images are available or not. Secondly, reanalysis datasets usually expand several decades enabling the generation of PV capacity factors for long-time periods. Thirdly, reanalysis dataset can be used to validate and bias correct climate models enabling the assessment of climate change impacts on energy system as in \cite{Kozarcanin_2018}. Fourthly, reanalysis datasets can also be used to generate time series representing the wind or hydroelectricity production and, hence, a consistent set of renewable generation time series to feed-in a certain energy model can be obtained. Finally, reanalysis data is usually freely available making it suitable for scientific analysis and the replicability of results. 

\

In summary, we know that the accuracy of reanalysis may not be sufficient for detailed studies on the performance of individual sites but we also know that there are significant benefits from using reanalysis to represent country-wise PV time series. Then, we can formulate the fundamental research questions of this paper as follows:
\begin{enumerate}
\item Can we use bias-corrected global reanalysis to obtain PV time series integrated over large-scale regions and attain a similar accuracy than when using satellite-based irradiance?
\item Can we develop a methodology to bias-correct reanalysis irradiance that is globally applicable?
\end{enumerate}

\

The Renewable Energy Atlas (REatlas) from Aarhus University was introduced in \cite{Andresen_2015} where it was used to obtain time series for onshore and offshore wind generation in Denmark that were validated against historical data. The REatlas uses as input the Climate Forecast System Reanalysis (CFSR) from the National Center for Environmental Prediction (NCEP) \cite{CFSR}.
In this paper, we present a methodology to obtain PV hourly capacity factors at a national level based on irradiance from reanalysis data. The method comprises two major innovations:

\begin{itemize}
\item[--] In the first place, we introduce a procedure to bias-correct irradiance at a country level using 12-values, one per month, which can be derived from the best available source, either satellite datasets or ground measurements. The bias-corrected irradiance is used, together with a model of the PV system, to generate hourly capacity factors for this technology at a national scale. These are validated against historical data. This allows us to retain the previously stated advantages of using irradiance from reanalysis datasets while reducing the errors due to a poor representation of the local atmosphere.
\item[--] In the second place, we propose an indirect procedure to infer the configuration (orientation and inclination) of PV panels in a country based on two artificial clear-sky days. By selecting historical PV electricity generation in hours with clear-sky conditions within days close to the summer and winter solstices we produce these two artificial clear-sky days corresponding to the days in which the sun is highest and lowest above the horizon. 
%We then assume that the orientation and inclination of PV panels in a country can be represented by two Gaussian distributions and select the adequate parameters of those distributions to replicate the two artificial clear-sky days. 
\end{itemize}

Finally, the bias-corrected irradiance is used to generate 38 years-long hourly PV time series for every country in Europe (EU-28 plus Serbia, Bosnia-Herzegovina, Norway, and Switzerland) and the modeled time series are validated using historical data for 2015 and 15 countries. The accuracy of the modeled time series is compared to that of EMHIRES \cite{EMHIRES} and Renewables Ninja \cite{Pfenninger_2016} datasets. The 38 years-long time series are computed for every European country and 4 possible configurations for the PV systems: (a) rooftop installations, (b) optimum orientation and inclination, (c) 2-axis tracking and (d) delta configuration. The time series are open-licensed and can be retrieved from the Zenodo repository. As a proof of concept, these time series are used to investigate the evolution of the mismatch curve, \textit{i.e.}, the electricity demand minus the PV generation, in two representative days (winter and summer solstice) in every country. 

\

The paper is organized as follows. Section \ref{data} summarizes all the data used. Methods are described in Sections \ref{methods} and \ref{sec_REatlas}. First, Section \ref{methods} describes the determination of monthly correction factors for reanalysis data. Then, a general overview of REatlas is provided on Section \ref{sec_REatlas} while annex A includes a detailed description of the model to convert irradiance into electricity generated by PV systems. The methodology to infer the orientation and tilt representative angles for every country based on two artificial clear-sky days is described in Section \ref{sec_determine_tilt}. Section \ref{results} compares the modeled time series with historical values provided by national TSOs. The analysis of mismatch curves is carried out in Section \ref{sec_duck_curves}. Finally, Section \ref{conclusions} gathers some conclusions. 
 
\begin{figure}[ht!]
\centering
\includegraphics[width=\columnwidth]{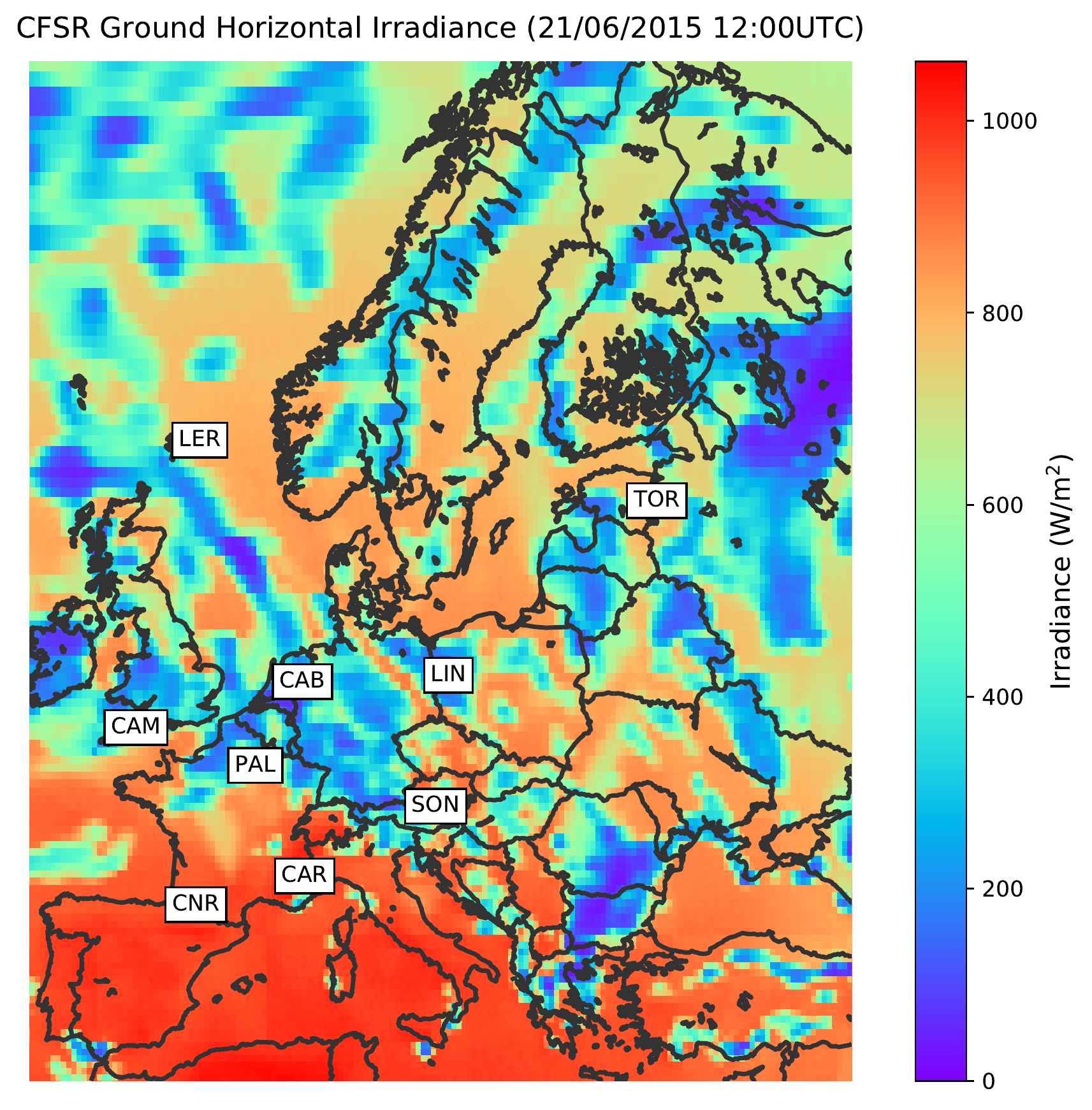}\\
\caption{Global Horizontal Irradiance from CFSR reanalysis dataset on the 21st of June, 2015 at 12:00 UTC. The location of every BSRN station corresponds to the lower-left corner of the 3-letter code labels shown in the map.} \label{fig_map}
\end{figure}

\section{DATA}
\label{data}

\subsection{Irradiance data}

\

\subsubsection*{Baseline Surface Radiation Network (BSRN)}
The Baseline Surface Radiation Network (BSRN) provides high temporal resolution measurements over long periods with high data quality. Global horizontal irradiance time series measured in the 9 ground stations located in Europe (Figure \ref{fig_map}) were downloaded and processed. The BSRN uses Secondary Standard ventilated pyranometers and the global measurement accuracy is estimated to be approximately 5 W/m$^2$ \cite{Ohmura_1998}. Besides the quality assurance protocols \cite{Konig_2013} implemented in the BSRN, we also applied the recommended ``Extreme Rare Limits'' and the comparison test to ensure the consistency of the global, direct and diffuse radiation measurements \cite{Long_2002}. Since BSRN provides minute-resolved measurements, a 60 minutes-wide averaging window has been applied to calculate hourly values that can be compared to those included in the CFSR reanalysis dataset. Assuming a horizontal wind speed of 10 m/s, the averaging window represents the movement of an air parcel of 36 km over the ground station. This can be compared to the irradiance data in reanalysis dataset whose spatial resolution is 40x40km$^2$. BSRN measurements are only used for the preliminary investigation on the capability of CFSR reanalysis dataset to capture local atmospheric effects carried out in Section \ref{PDF_clearness_index} but this information is not used in the methodology to determine the monthly correction factors proposed in this paper. 

\

%Modern-Era Retrospective Analysis for Research and Applications-2 (MERRA-2) is a global reanalysis produced by the NASA Global Modeling and Assimilation Office (GMAO, [REF MOLOD]). It provides data since 1980 at hourly resolution and spatial resolution equal to 0.5$^{\circ}$ latitude and 0.625$^{\circ}$ longitude, that in Europe translates to approximately 50x50km$^2$. On the contrary to its predecessor MERRA, MERRA-2 includes space-based aerosol observation but as shown in \cite{Pfenninger_2016} the accuracy when modeling PV is not improved. The global horizontal irradiance at ground level $G(0,t)$ and the extraterrestrial irradiance $B(t)$ for the locations of the BSRN ground stations within Europe was downloaded from the renewables ninja webpage \cite{ninja}. It is known that MERRA tends to overestimate the total irradiation because it frequently underestimates the presence of clouds \cite{Boilley_2015, Pfenninger_2016}. 
\subsubsection*{Climate Forecast System Reanalysis}
The Climate Forecast System Reanalysis (CFSR) is provided by the National Center for Environmental Prediction (NCEP) \cite{CFSR}. It comprises a 38 years-long global high-resolution dataset (hourly time resolution and spatial resolution of 0.3125$^{\circ}$ x 0.3125$^{\circ}$ which in Europe is roughly equivalent to 40 x 40 km$^2$). The irradiance data included in the CFSR dataset is used as input for the REatlas to generate PV capacity factors time series. Prior to the conversion, the monthly bias correction described in Section \ref{monthly_correction_factors} is carried out. Figure \ref{fig_map} shows the irradiance from CFSR reanalysis dataset on the 21$^{st}$ of June, 2015 at 12:00 UTC

\
\subsubsection*{Solar Surface Radiation - Heliostat (SARAH)}
The Meteosat-based SARAH (Solar SurfAce RAdiation - Heliostat) satellite dataset provides hourly resolution and very high spatial resolution (0.05$^{\circ}$x0.05$^{\circ}$) for 30 years (1986-2015) although a significant percentage of the values are missing for the initial years \cite{Pfenninger_2016}. The SARAH irradiance dataset has been used to retrieve ground horizontal irradiance time series for every location corresponding to a point in the CFSR grid. Those time series are aggregated to generate irradiance time series at country level that are compared to those generated using CFSR reanalysis to determine monthly correction factors. The procedure is described in Section \ref{monthly_correction_factors}. The SARAH irradiance time series for different locations are downloaded using PV-GIS version 5 \cite{PVGIS}. This is also ensures the replicability of the method. For additional countries to those included in this paper, the best available irradiance data can be used to correct CFSR reanalysis irradiance on a monthly basis. For instance, the National Surface Radiation Database (NSRDB) maintained by NREL \cite{NSRDB} can be used to determine correction coefficients in North and South America. 
%ERA-5 satellite dataset can be used to determine monthly correction coefficients in northern European countries whose area is not totally included within SARAH datasets or 

\subsection{PV cumulative installed capacity}
\begin{figure*}[ht!]
\centering
\includegraphics[width=0.9\textwidth]{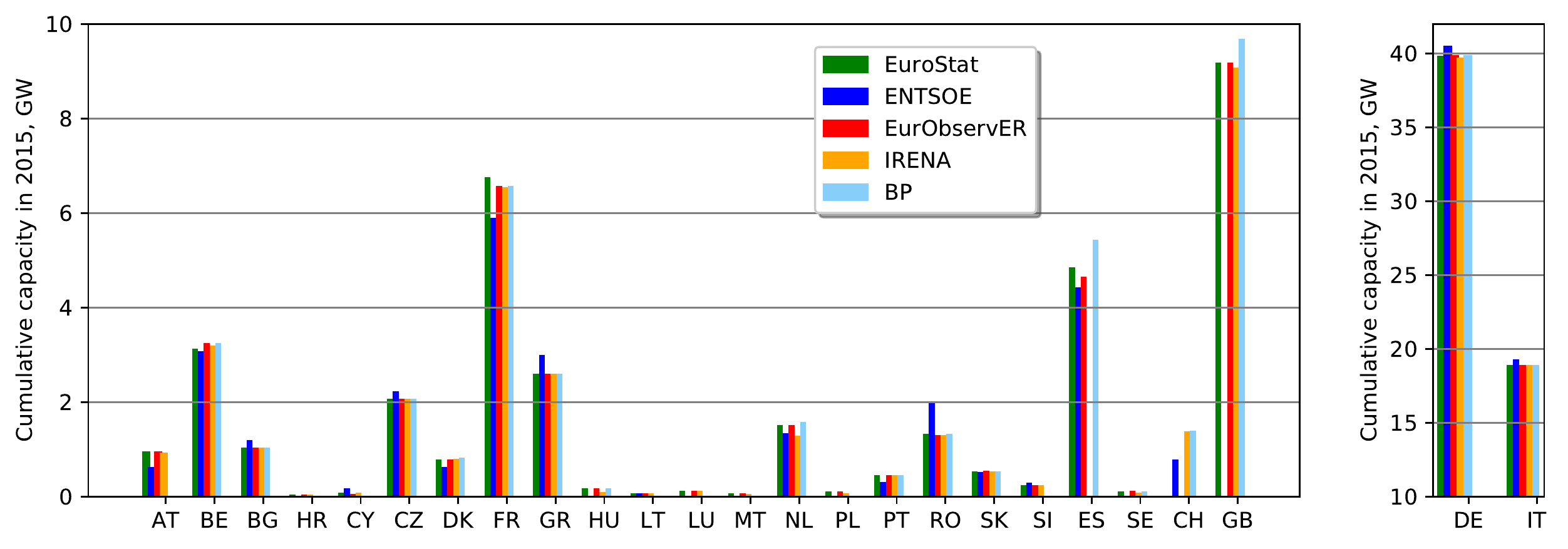}\\
\caption{Cumulative installed capacities for European countries in 2015 reported by different sources.} \label{fig_PV_capacities}
\end{figure*}

The cumulative installed capacities for every European country in 2015 were obtained from the following sources: ENTSO-E, Eurostat, EurObservER \cite{EurObservER_2017}, IRENA \cite{IRENA_2018}, BP \cite{BP}. The data from the first two sources were retrieved through the convenient compilation carried out by the Open Power System Data (OPSD) initiative \cite{OPSD_capacities}. Figure \ref{fig_PV_capacities} shows the installed capacities according to various sources. The large discrepancy found is probably a combined effect of a fast changing scenario, where significant capacities relative to cumulative values are installed every year, and the difficulties associated to monitoring a myriad of new small installations. Since the discrepancy between different sources is significant, an averaged value was calculated and used to compute the historical hourly capacity factors in the next section. This is the same approach followed in \cite{Pfenninger_2016}. Cumulative installed capacities are considered to be constant throughout 2015, since differences among the capacities reported by different sources are in some cases larger than the difference among two consecutive years reported by the same source.

\subsection{PV generation time series and hourly capacity factors} \label{sec_TSO_data}

Actual PV generation time series are reported by ENTSO-E or national TSOs for 2015 for the following countries: Austria, Belgium, Bulgaria, Czech Republic, Denmark, France, Germany, Greece, Italy, Lithuania, Netherlands, Portugal, Romania, Slovakia, Slovenia and Spain. The data is accessed through the OPSD file \cite{OPSD_timeseries}. For the case of Spain, ENTSO-E data cannot be used since only the solar aggregated time series, including the generation from Concentrated Solar Power (CSP) and PV plants, is reported. Hence, hourly values for every day starting from May, 1$^{st}$, 2015 have been retrieved from \cite{REE_web} and assembled. The hourly capacity factors for every country in 2015 are computed by dividing the PV hourly electricity generation by the cumulative installed capacity, averaged values among those reported by different sources.
%Additionally, data for PV generation in Italy and Spain corresponding to 2016 (the first year when it was reported) is used.   

\section{METHODS: Determination of country-wise monthly correction factors for reanalysis irradiance }
\label{methods}

\subsection{Preliminary analysis: reanalysis capability to capture local atmospheric effects} \label{PDF_clearness_index}

\begin{figure}[ht!]
\centering
\includegraphics[width=0.8\columnwidth]{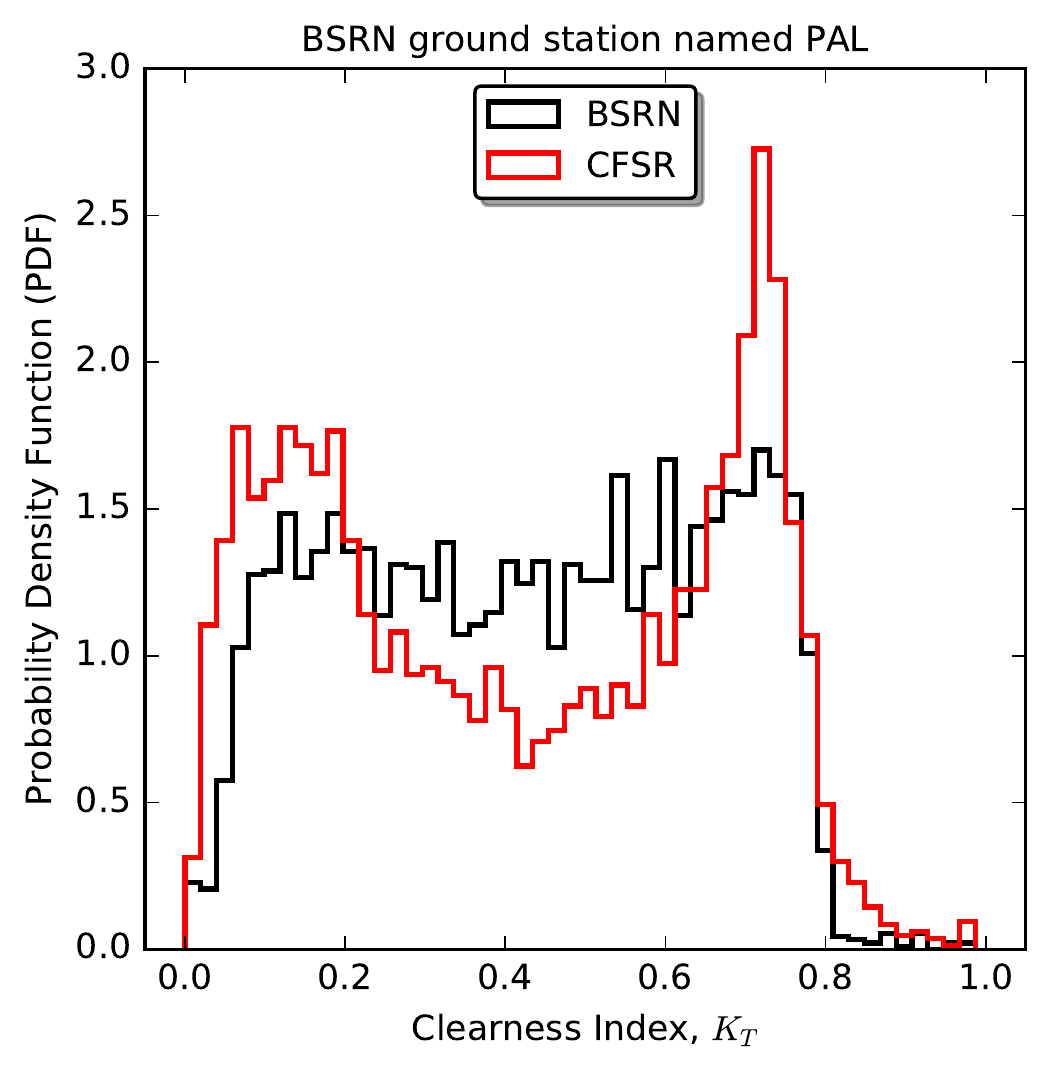}\\
\caption{Probability Density Function (PDF) of the clearness index $K_t$ calculated from measurements at the BSRN station located in Palaiseau, France and from the CFSR reanalysis dataset. The bin sizes are 0.02.} \label{fig_PDF_clearness_index}
\end{figure}

The clearness index $K_t$ is defined as the ratio between the global irradiance at the ground $G(0)$ and the extraterrestrial irradiance $B_0$, that is, at the top of the atmosphere (equation \ref{eq_clearness_index}). $K_t$ is influenced by the thickness of the atmosphere, which in turns depends on the time, date, and location, as well as by its composition and cloud content. Furthermore, $K_t$ is usually employed to calculate the fraction of direct to global irradiance at ground level (equation \ref{eq_F} and \cite{Lorenzo_Handbook}). We follow the approach proposed in \cite{Frank_2018} to evaluate the capability of the CFSR reanalysis dataset to represent the local atmosphere filtering properties. In Figure \ref{fig_PDF_clearness_index}, the probability density function (PDF) of $K_t$ for every hour throughout 2015 is depicted for the BSRN ground station located in Palaiseau, France. The figure shows the PDF obtained from the time series corresponding to irradiance at the CFSR grid point closest to the station together with the PDF obtained from ground measurements. When compared to experimental data, the CFSR PDF shows a higher probability for very-low and very-high clearness indices. The same result is consistently found for the 9 ground stations within the BSRN located in Europe. The associated figures are provided in the Supplementary Materials. The impact of the atmosphere in the CFSR dataset is more extreme than in reality. For clear-sky days, the modeled atmosphere is more transparent than in measurements overestimating global horizontal irradiance, while, on cloudy days, the modeled atmosphere is more absorbing/scattering than in reality underestimating global horizontal irradiance.

\subsection{Monthly correction factors} \label{monthly_correction_factors}

The preliminary analysis demonstrated that the CFSR reanalysis dataset does not properly represent the light filtering effect of the local atmosphere. Since PV generation is directly proportional to the irradiance, any bias in irradiance must be corrected to avoid a significant error when computing PV time series. A classic approach followed to estimate the energy produced by a PV power plant consists in using monthly values for $K_t$ to characterize the solar climate at a particular location. By using one $K_t$ value per month, the dispersion is reduced and a better match with the real performance of the power plant in achieved \cite{Lorenzo_Handbook}. Based on that experience, we propose the following procedure to bias correct irradiance values from CFSR.
\begin{enumerate}
\item For every country, global horizontal irradiance time series $G^{CFSR}_n(0,t)$ for every CFSR grid point $n$ within a country are retrieved and aggregated to obtain a time series $G^{CFSR}_s(0,t)$ representative for the country $s$.

\begin{equation}
G^{CFSR}_s(0,t) = \sum G^{CFSR}_n(0,t)  
\end{equation}
where $n \in s$.

\item Global horizontal irradiance time series for the same locations $G^{SARAH}_n(0,t)$ are obtained from SARAH dataset and aggregated to obtain $G^{SARAH}_s(0,t)$. 

\begin{equation}
G^{SARAH}_s(0,t) = \sum G^{SARAH}_n(0,t)  
\end{equation}
where $n \in s$. Satellite-based SARAH dataset was selected since it is probably the most accurate dataset available \cite{Urraca_2017}. Alternatively, for countries not included in SARAH dataset, another satellite dataset or measurements from a ground stations network can be used.

\item	For every month $m$, the correction factor $C_{m,s}$ is estimated as that of the median day, that is
\begin{equation}
C_{m,s}=median\left( \frac{\int_{day} G^{SARAH}_s(0,t) {d}t}{ \int_{day} G^{CFSR}_s(0,t) {d}t}  \right) 
\end{equation}
where $day \in m$.
\item For every country, $C_{m,s}$ are determined using irradiance data for years in the period 2005-2014 and the average is calculated. Then, $C_{m,s}$ are used to correct CFSR time series modeled for 2015. 

\begin{equation}
G^{CFSR, corrected}_s(0,t) =C_{m,s} G^{CFSR, uncorrected}_s(0,t) 
\end{equation}
where $t \in m$.
\end{enumerate}

\begin{figure}[ht!]
\centering
\includegraphics[width=\columnwidth]{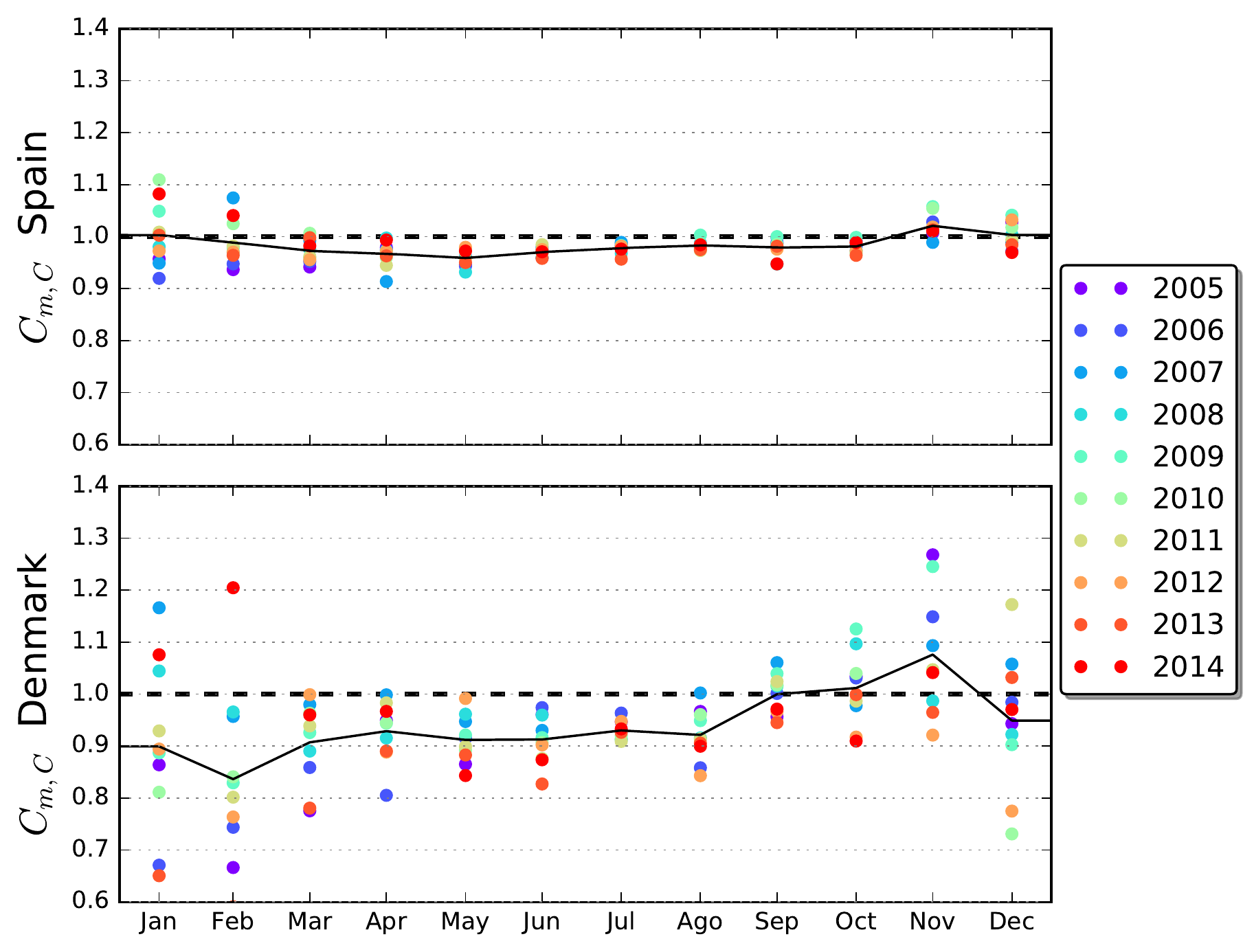}\\
\caption{Monthly correction factors $C_{m,s}$ for Spain and Denmark. Figures for other European countries are included in the Supplementary Materials.} \label{fig_monthly_2countries}
\end{figure}

The correction factors $C_{m,s}$ represent the solar climate characteristic of a country and correct the irradiance from CFSR in the cases where the reanalysis is not capable of adequately capturing it.  In general, irradiance values in the CFSR dataset are higher than those obtained from the SARAH dataset and, consequently, monthly correction factors are lower than 1. This was expected as other reanalysis datasets are known to underestimate the presence of clouds \cite{Boilley_2015, Urraca_2017}. This is also in agreement with the Europe-wide scaling factor equal to 0.935 applied to correct the Renewables Ninja time series obtained using irradiance from the MERRA-2 reanalysis dataset \cite{Pfenninger_2016}. $C_{m,s}$ show noticeable differences among countries. For instance, $C_{m,s}$ determined for Spain depicted in Figure \ref{fig_monthly_2countries} are very close to one for every month. Conversely, monthly correction factors for Denmark varies throughout the year and are higher for winter months. The large variability in $C_{m,s}$ winter values for Denmark is caused by the unstable and cloudy local climate during those months. $C_{m,s}$ values are summarized in \ref{annex_monthly_correction_factors} and figures for every European country are provided in the Supplementary Materials. 

\section{METHODS: Computation of PV hourly capacity factors at country level using REatlas} \label{sec_REatlas}

The REatlas from Aarhus University is used for converting irradiance and temperature time series into country-wise PV hourly capacity factors. The REatlas was introduced in \cite{Andresen_2015} and a detailed description of the methodology and equations involved in the solar conversion can be found in \ref{annex_solar_conversion}.  The procedure can be summarized as follows. For every point in the CFSR grid, the irradiance time series is bias corrected as described in Section \ref{monthly_correction_factors}. The global irradiance at ground level is first decomposed into direct and diffuse irradiance (eq. \ref{eq_F}). Then, the direct, diffuse, and global irradiances on a tilted panel are calculated and aggregated (eq. \ref{eq_G_B_D}). The global irradiance at the entrance of the solar panel is converted into electricity using a simplified model for the PV system. The PV output is assumed to be proportional to the irradiance at its entrance. The impact of temperature on efficiency is assessed using the classic approach based on the Nominal Operating Cell Temperature (NOCT) (eq. \ref{eq_TONC} and \cite{Lorenzo_Handbook} ). The efficiency temperature coefficient of crystalline silicon flat panel is assumed since this is the most spread technology. Finally, the time series for every point in the CFSR grid data within a country are aggregate to obtain hourly capacity factors representative for every country. A uniform capacity distribution across every country is assumed. In practice, this implies assuming one PV panel is installed in every point in the CFSR grid. Although detailed databases including the location of wind turbines \cite{thewindpower} and conventional power plants exists, this is not the case of PV plants. The uncertainty associated to the real position of thousands of small PV installations lead us to select the uniform distribution hypothesis.

\section{METHODS: Determination of the configuration of PV panels based on artificial clear-sky days.} \label{sec_determine_tilt}

\begin{figure}[ht!]
\centering
\includegraphics[width=0.8\columnwidth]{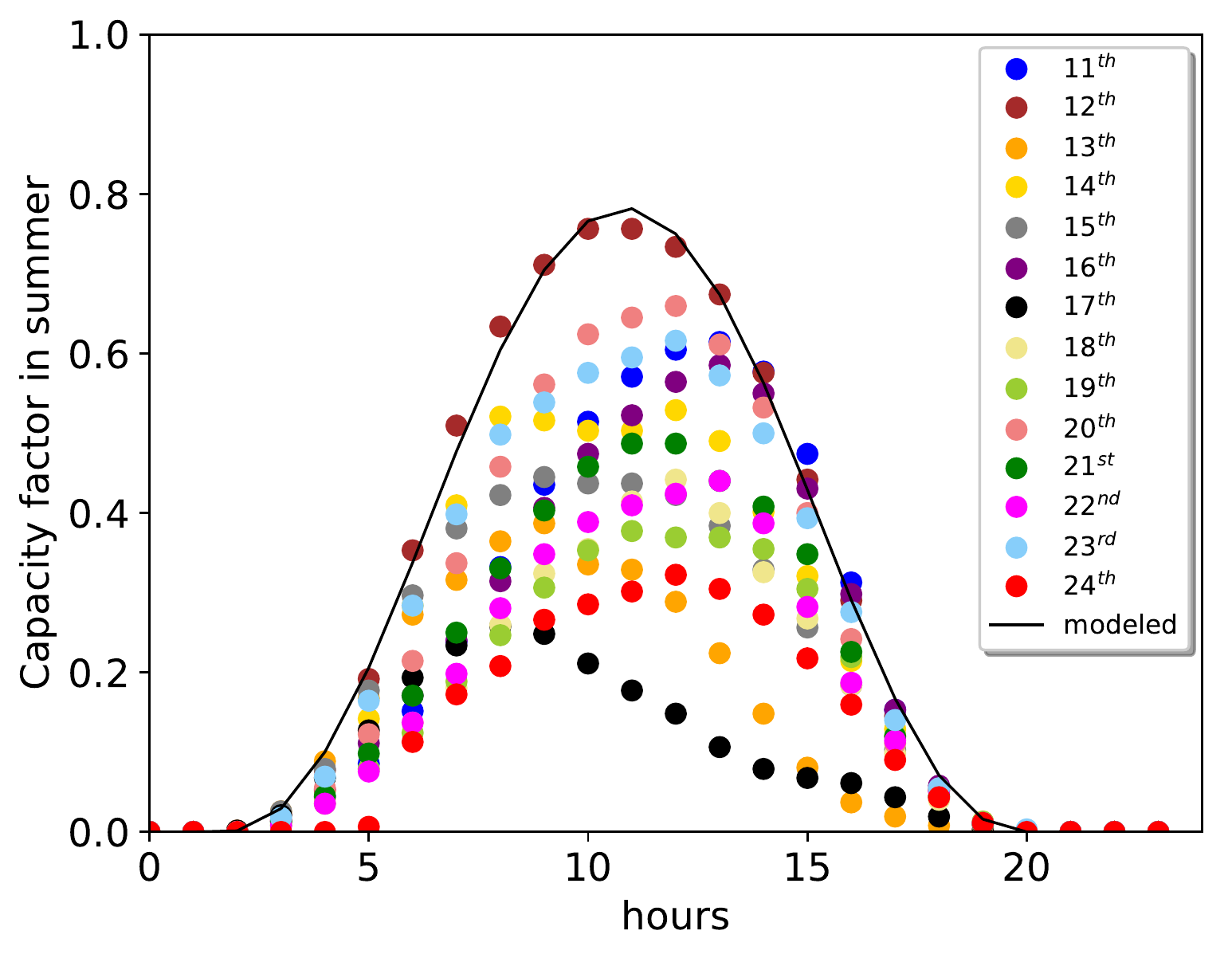}\\
\caption{Capacity factor throughout the artificial clear-sky summer solstice in Denmark obtained by selecting, for every hour, the maximum capacity factors for days in June close to the 21$^{st}$.} \label{fig_clear_sky_DNK}
\end{figure}

We assume that the configuration of PV panels in a country can be represented using two normal distributions, that is, the tilt angles are assumed to follow a Gaussian distribution centered in $\mu_{\beta}$ with a standard deviation $\sigma_{\beta}$ while the orientation angles follow a Gaussian distribution centered in $\mu_{\alpha}=0^{\circ}$, on average PV panels are south oriented, and with a standard deviation $\sigma_{\alpha}$. In principle, if winter and summer solstices happen to be clear-sky days, the parameters $\sigma_{\beta}$, $\mu_{\alpha}$, and $\sigma_{\alpha}$ could be inferred by comparing the modeled hourly capacity factors at national level throughout those days to those reported by the TSOs. Under the clear-sky assumption, differences between the PV generation in winter and summer solstices are directly related to the inclination and orientation of PV panels. The main problem is, of course, that clear-sky conditions may not occur in winter and summer solstices. In fact, since we are aiming to uniform clear-sky conditions across the whole country, it is possible that there is not a single clear-sky day throughout the year. However, we can take advantage of the fact that the Sun path across the sky is almost constant for the days close to the solstice. We have generated two artificial clear-sky days by selecting, for every hour, the maximum capacity factor found for the days around the summer and winter solstice. For instance, Figure \ref{fig_clear_sky_DNK} depicts the artificial clear-sky summer solstice reconstructed for Denmark. We have selected a $\pm$10 days window around the 21$^st$ of June and the 21$^st$ of December. Within those days and for the latitudes in Europe, the maximum daily irradiance varies within 5\% of the value found in the solstice. A figure showing this variation for different countries is provided in the Supplementary Materials.

The theorical hourly capacity factor for winter and summer solstices is calculated as

\begin{equation} \label{eq_ideal_CF}
CF_{s} (t) =\sum_{n, \beta, \alpha} \zeta CF_n(t,\beta,\alpha) PDF(\mu_{\beta},\sigma_{\beta}) PDF(\mu_{\alpha},\sigma_{\alpha}) 
\end{equation}

where the parameter $\zeta$ represents the decrease in the maximum capacity factor due to one of the following causes: (a) the time evolution of irradiance in different location is asynchronous due to different latitudes and longitudes in a country, (b) presence of clouds in some part of the country, (c) systems out of production due to repairing or maintenance, (d) systems under curtailment following TSO orders. 
The parameters $\zeta$ and $\mu_{\beta}$ are inferred by minimizing the normalized Root Mean Square Error ($RMSE_{n}$) calculated by comparing the modeled capacity and the artificial clear-sky winter and summer solstices recreated from historical PV generation reported by TSO (and assuming $\mu_{\alpha}$=0$^{\circ}$, $\sigma_{\beta}$=20$^{\circ}$, and $\sigma_{\alpha}$=30$^{\circ}$). \ref{annex_gaussian_parameters} gathers the optimum parameters estimated for different countries. We did not find a direct correlation between $\mu_{\beta}$ and the latitude representative for every country (Figure is provided in the Supplementary Materials). This indicates that the panel tilt angles are probably more influenced by the rooftop inclination than by the local latitude. This is in agreement with Pfenninger and Staffel \cite{Pfenninger_2016} findings by processing metadata of individual sites. Consequently, for the system configuration named as rooftop in section \ref{results}, we assumed the same values for every country ($\zeta=1$, $\mu_{\alpha}$=0$^{\circ}$, $\mu_{\beta}$=25$^{\circ}$, $\sigma_{\alpha}$=40$^{\circ}$, and $\sigma_{\beta}$=15$^{\circ}$).

%\textcolor[rgb]{1,0,0}{Comentar al final las características de esta tabla. En todos ellos las sigmas son grandes. En la mayoría tilt medio es de 20-30º, esto coincide con lo que obtiene ninja a partir del procesado de los metadatos de un montón de instalaciones individuales. }

\section{RESULTS: Modeled vs historical time series}
\label{results}

\begin{figure*}[ht!]
\centering
\includegraphics[width=0.9\textwidth]{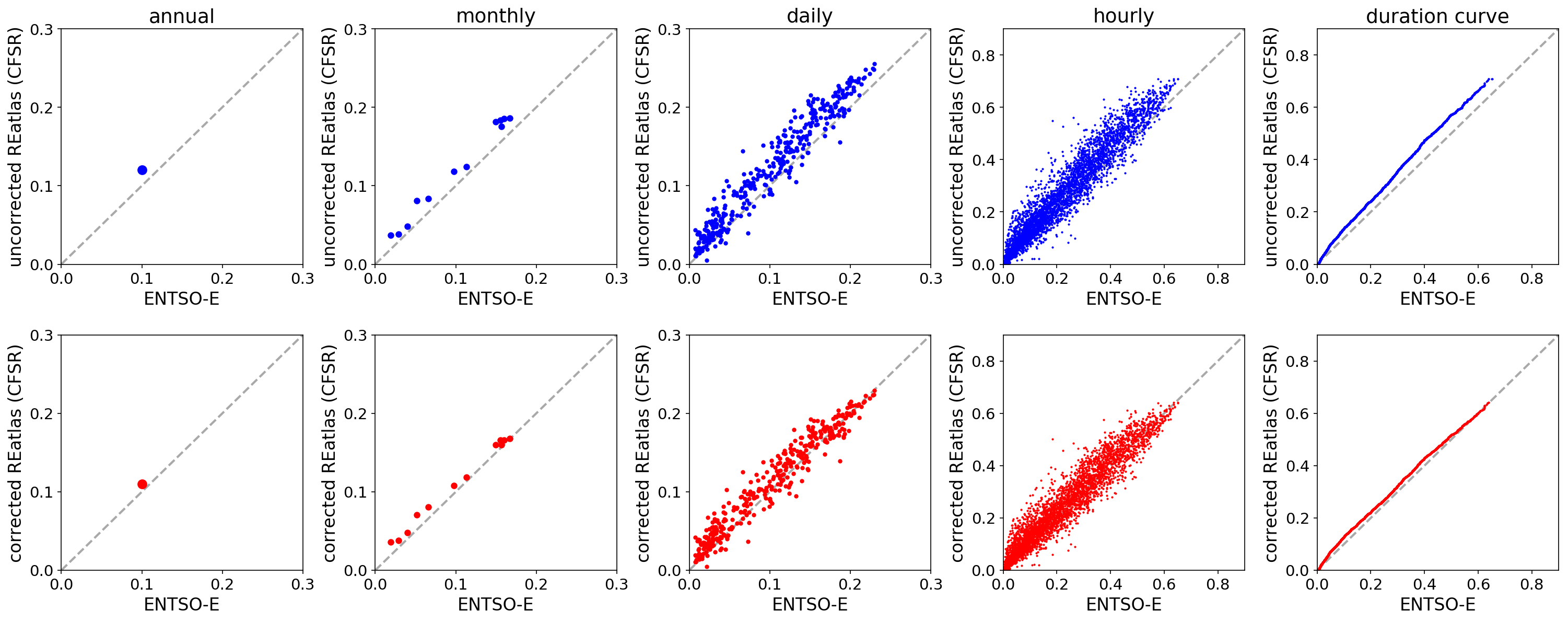}\\
\caption{QQ plots comparing modeled and historical capacity factors for Germany integrated throughout four different time periods (from left to right: year, month, day, hour). The rightest column shows the QQ plot for the duration curve (sorted hourly capacity factors).} \label{fig_QQ_plot}
\end{figure*}

\begin{figure*}[ht!]
\centering
\includegraphics[width=0.8\textwidth]{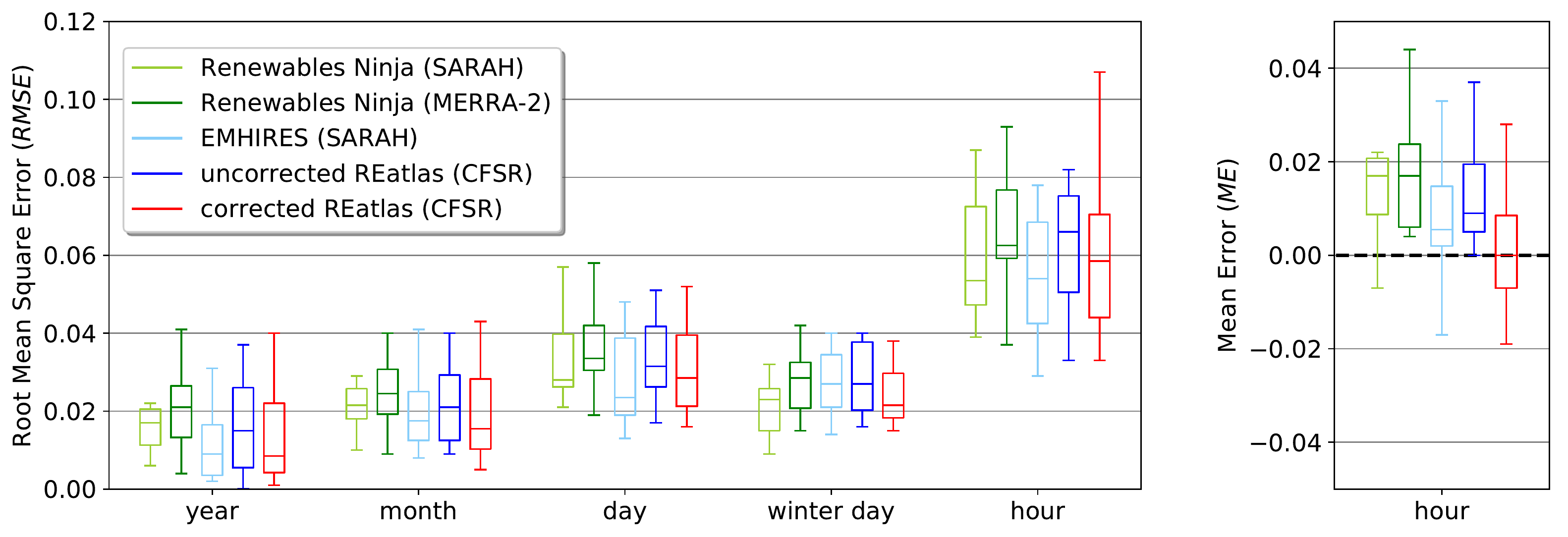}\\
\caption{(left) Root Mean Square Error ($RMSE$, eq. \ref{eq_RMSE}) and Mean Error ($ME$, eq. \ref{eq_ME}) calculated comparing simulated and historical time series for 2015 and 15 countries. } \label{fig_boxplot}
\end{figure*}

The country-wise capacity factors modeled using REatlas and bias-corrected reanalysis irradiance for 2015 are compared to historical values. Figure \ref{fig_QQ_plot} depicts the QQ plots for Germany when the capacity factors are integrated using different time scales (year, month, day, and hour) as well as the QQ plot for the duration curve (sorted hourly capacity factors). Figures for other countries are provided in the Supplementary Materials. To enable a global assessment of the modeled time series, Figure \ref{fig_boxplot} depicts the RMSE and Mean Error (ME) calculated including every country $s$ where historical data is available (see Section \ref{sec_TSO_data}) and using different time periods $p$. The $RMSE_p$ and $ME_p$ are defined as follows:

\begin{equation} \label{eq_RMSE}
RMSE_{p}= \sqrt{\frac{\sum_{s,p} (\int_p CF_s^{mod}(t) {d}t - \int_p CF_s^{hist}(t) {d}t)^2}{n_s \cdot n_t}}
\end{equation}

\begin{equation} \label{eq_ME}
ME_{p}=\frac{\sum_{s,p} (\int_p CF_s^{mod}(t) {d}t - \int_p CF_s^{hist}(t) {d}t)}{n_s \cdot n_t}
\end{equation}

\begin{figure*}[ht!]
\centering
\includegraphics[width=0.75\textwidth]{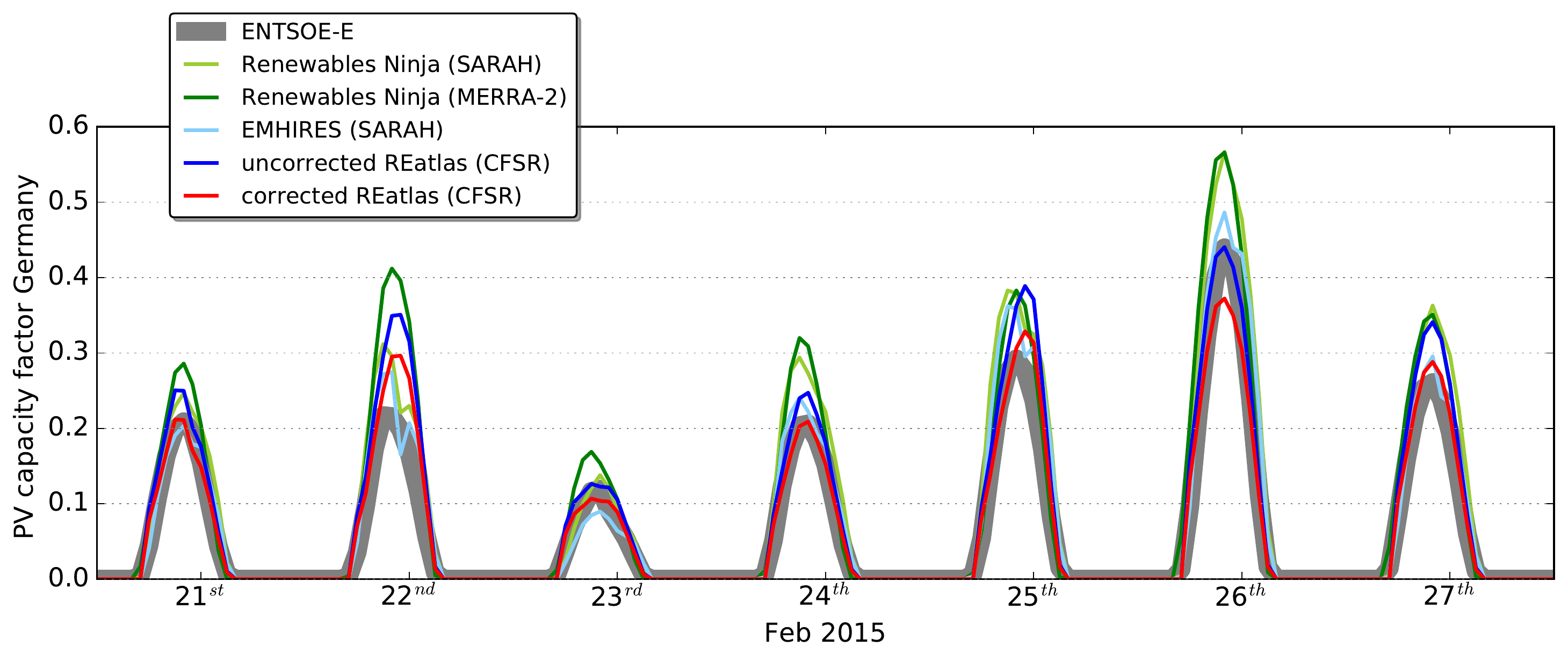}\\
\caption{Modeled time series vs historical data throughout a week in February for Germany.} \label{fig_winter_week}
\end{figure*}

\begin{figure*}[ht!]
\centering
\includegraphics[width=0.75\textwidth]{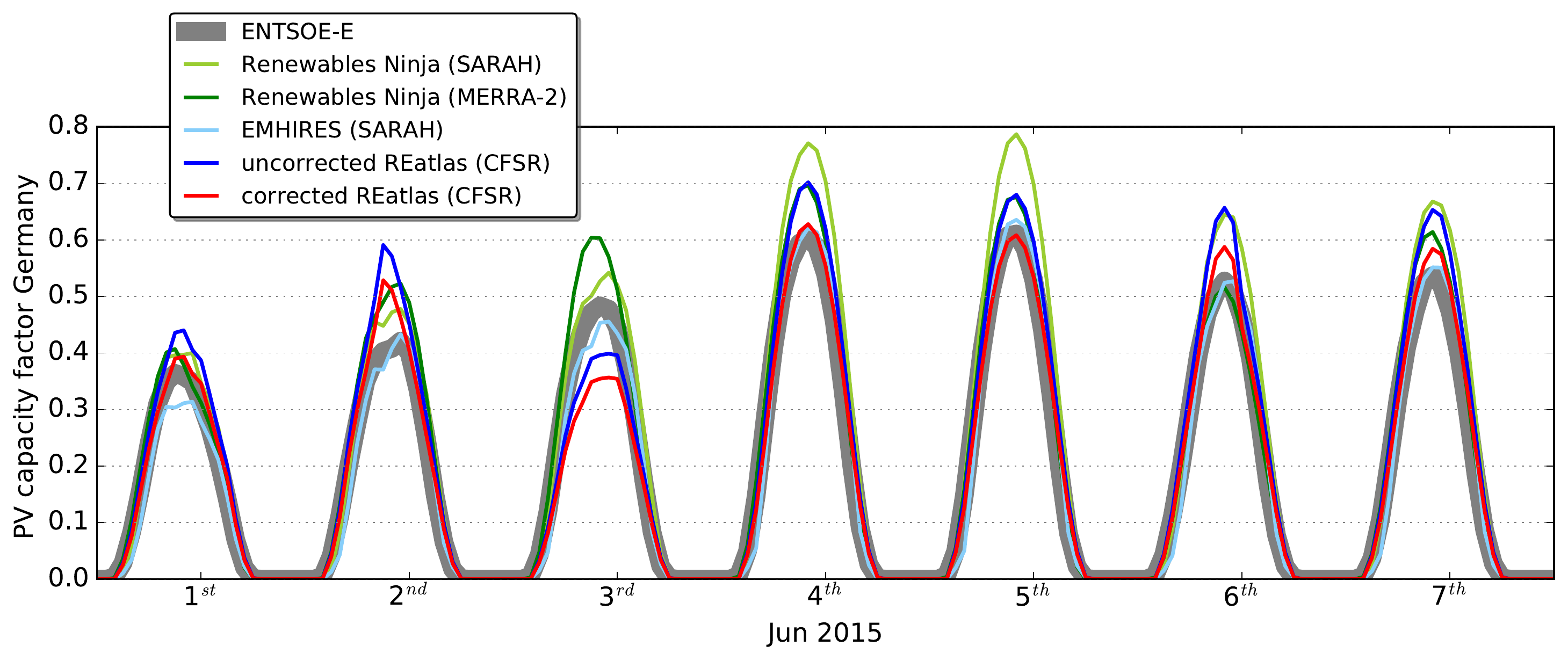}\\
\caption{Modeled time series vs historical data throughout a week in June for Germany.} \label{fig_summer_week}
\end{figure*}

where $n_s$ is the number of countries and $n_p$ the number of periods. Figure \ref{fig_boxplot} also includes the errors calculated using time series for 2015 provided by Renewables Ninja (using either SARAH or MERRA-2 dataset as input) \cite{Pfenninger_2016} and the EMHIRES time series (using SARAH dataset as input) \cite{EMHIRES}. Figures \ref{fig_winter_week} and \ref{fig_summer_week} depict, for Germany, the time evolution of PV throughout two representative weeks, modeled by different sources together with the historical data provided by the TSO. From Figures \ref{fig_QQ_plot} and \ref{fig_boxplot} it can be clearly observed that, for all datasets, the $RMSE$ increases as we try to model capacity factors integrated over shorter time periods. This is a classic result when modeling the generation of PV plants: the uncertainty in the prediction of the energy generated by the plant in a certain month is far lower than the uncertainty when attempting to predict the generation in a particular day of that month \cite{Lorenzo_Handbook}. It is also one of the  main reasons that led us to propose monthly correction factors for the irradiance.

\

Although most of the energy models use hourly values, it is important to realize that uncertainty in the energy model outcomes may be influenced by a different timescale. For example, let's imagine that we try to optimize the capacity mix necessary to supply the electricity demand in a country using mainly renewables and that it is cost beneficial to provide a significant share of the electricity using PV. In that case, several works \cite{Schlachtberger_2017, Brown_2018, Rasmussen_2012} have found that electric batteries must be installed to counterbalance the daily cycles of PV generation and that the necessary energy and power capacity of those batteries are heavily impacted by the PV generation throughout winter weeks with low renewable generation. Then, in order to reduce the uncertainty of the model results, it becomes more important to properly simulate the daily PV generation throughout those weeks than to match exactly the hourly generation. 

\

Figure \ref{fig_boxplot} shows how the irradiance-corrected REatlas time series adequately predict the country-wise performance of PV at different time scales. The RMSE calculated for the irradiance-corrected REatlas time series show, for all the different integration periods, similar RMSE to Renewables Ninja time series (either using MERRA-2 or SARAH dataset as input) and EMHIRES. It should be reminded here that the monthly correction factors for the irradiance time series have been calculated using irradiance from the previous years (2005-2014). However, the EMHIRES time series have been corrected using the PV generation reported from TSO in 2015, that is, the same historical data against which they are evaluated, so a very good match was expected.  For the Renewables Ninja time series, Figure \ref{fig_boxplot} confirms the lower RMSE when using SARAH dataset as input found for 2014 in \cite{Pfenninger_2016}.

\

Figure \ref{fig_boxplot} also shows the Mean Error ($ME$) calculated for all the hours in 2015 and the 15 countries with available historical data. The uncorrected REatlas time series shows a positive bias of 0.011 that is reduced to 0.001 for the irradiance-corrected REatlas time series. For 2016, the bias is 0.013 and 0.003 for corrected/uncorrected REatlas time series. For 2017, the attained values are 0.016 and 0.006 respectively. Equivalent figures using data for 2016 and 2017 are provided in the Supplementary Materials. When observing Figure \ref{fig_boxplot}, one meaningful question arises: Why the SARAH-based Renewables Ninja time series shows a positive ME while the corrected REatlas time series, whose bias correction is based on SARAH database, doesn't? We discuss bellow several possible explanations. First, SARAH-based Renewables Ninja time series were bias corrected using a Europe-wide scaling factor of 1.094. On the one hand, this factor was determined to minimize the mean bias when comparing the modeled time series with historical data in more than 1000 individual sites across Europe. On the other hand, the $ME$ for SARAH-based Renewables Ninja time series is computed by comparing the modeled time series with the historical data reported by TSOs. We agree with Pfenninger and Staffel on their affirmation that \textit{``the output reported by the TNOs is not necessarily more accurate than simulations''} so the $ME$ shown in Figure \ref{fig_boxplot} may not be the best metrics to assess the time series. Nevertheless, this is the most accessible historical data at these moments. There are two alternative explanations to this discrepancy. First, REatlas uses a composed circumsolar and isotropic model to represent diffuse irradiance on solar panels (equation \ref{eq_iso_circum}) while Renewables Ninja uses a isotropic-sky model. The latter is know to underestimate the diffuse irradiance on PV panels tilted to the equator, that could have led to the positive bias-correction used in Renewables Ninja time series. Second, SARAH is probably the most accurate available irradiance dataset and it has been validated against a dense network of ground stations showing a extremely low mean error for the whole Europe \cite{Urraca_2017}. However, it is know to slightly underestimate irradiance at high latitudes, overestimate it in the south and attain unbiased estimation in central Europe. Since the individual PV sites used in Renewables Ninja are not uniformly distributed across Europe, this could have influenced the calculation of the scaling factor. 

\begin{figure}[ht!]
\centering
\includegraphics[width=0.8\columnwidth]{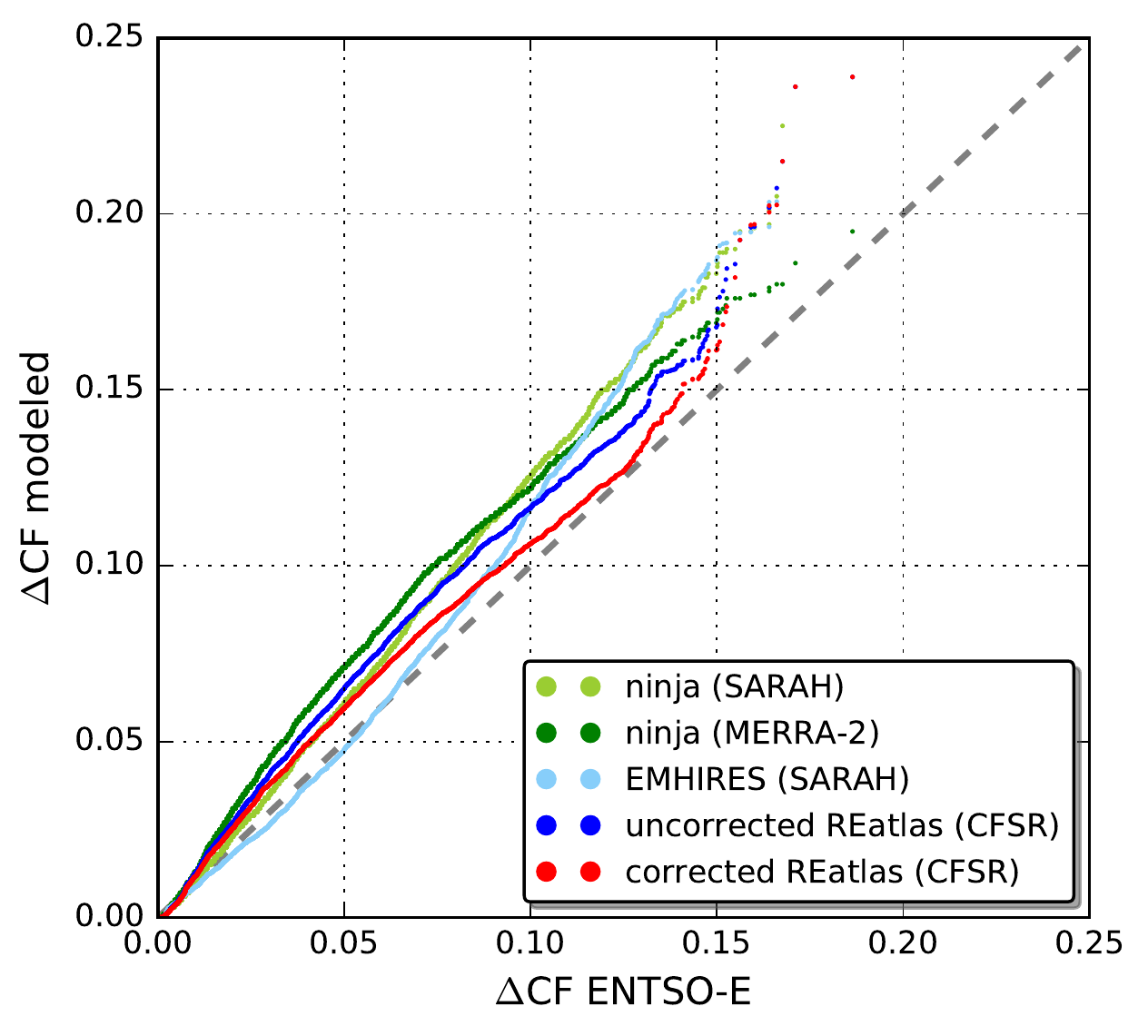}\\
\caption{QQ plot for the duration curve of the ramps $\Delta CF$ in Germany.}\label{fig_ramps}
\end{figure}
\

The distribution of ramps, \textit{i.e.}, the temporal variability of the hourly capacity factor, is a usual metric to evaluate modeled time series for wind energy \cite{Andresen_2015, Staffell_2016}. Extreme ramps in the PV generation time series might be critical for grid stability and require a fast response from backup technologies. Modeling the dynamics of ramps is easier for PV time series since they are mainly determined by the Sun daily cycle. Figure \ref{fig_ramps} depicts, for Germany, the QQ plots for the duration curve of the modeled and historical ramps. The hourly ramp $\Delta CF$ is defined as the difference between the capacity factor $CF(h)$ and its value in a previous hour $CF(h-1)$:
\begin{equation}
\Delta CF =\left. CF (h) - CF (h-1) \right.
\end{equation}

Equivalent figures for other countries are provided in the Supplementary Materials. In general, the ramp rates modeled using reanalysis irradiance are statistically consistent with the observed ones. Maximum ramps of $\Delta CF \approx$ 0.25 are found for Germany, this implies that the power supplied by PV increases or decreases by 25\% of the installed capacity in an hour. Nevertheless, it should be mentioned here that ramps are caused by two different phenomena. On the one hand, the Sun trajectory in the sky which can be accurately predicted. On the other hand, the country-integrated effect of clouds. 

\section{RESULTS: Impact of the PV system configuration on the mismatch curves} \label{sec_duck_curves}
 
Future PV capacity factors at country level are difficult to predict. For wind energy, higher capacity factors are expected for future years as wind turbines increase their size and efficiency \cite{Andresen_2015, Staffell_2016}. In particular, the deployment of offshore wind turbines will rise wind hourly capacity factors when aggregated at a country level. 
Conversely, PV capacity factors mainly depend on the system configuration (static vs. tracking, orientation and tilt angles). In fact, if the cost decrement tendency keeps as it is today, country-wise annual capacity factors will probably decrease. With very cheap PV panels, static configuration will be preferred, as the extra energy provided by the tracker will not pay off, PV will be integrated into building with non-optimum tilt and orientation angles and even delta configuration (a row of tilted panels where those on one side face east while those on the other face east) may result cost effective. 

\

Using bias-corrected reanalysis irradiance and the methodology described in Section \ref{sec_REatlas}, we have generated 38 years-long time series representing the PV hourly capacity factors in every country in Europe (EU-28 plus Serbia, Bosnia-Herzegovina, Norway, and Switzerland). A uniform capacity layout is used, that is, one PV panel is assumed to be installed in every point in the CFSR grid data, every 40x40 km$^2$, and the configurations of the panels is one of the followings:

\begin{enumerate}
\item \textbf{Rooftop} installation where tilt angles follow a Gaussian distribution with $\mu_{\beta}=25^{\circ}$ and $\sigma_{\beta}=15^{\circ}$ and orientation angles follows a Gaussian distribution centered around 0$^{\circ}$ (south orientation) with $\sigma_{\alpha}=40^{\circ}$. These are the same distributions assumed in \cite{Pfenninger_2016}.
\item South-orientation ($\alpha$=0$^{\circ}$) and \textbf{optimum tilt} angle for every panel.
\item 2-axis \textbf{tracking}.
\item \textbf{Delta} configuration where tilt angle is 30$^{\circ}$, one panel is oriented to the east ($\alpha$=90$^{\circ}$) and the other is  is oriented to the west ($\alpha$=-90$^{\circ}$). 
\end{enumerate}

\begin{figure}[ht!]
\centering
\includegraphics[width=\columnwidth]{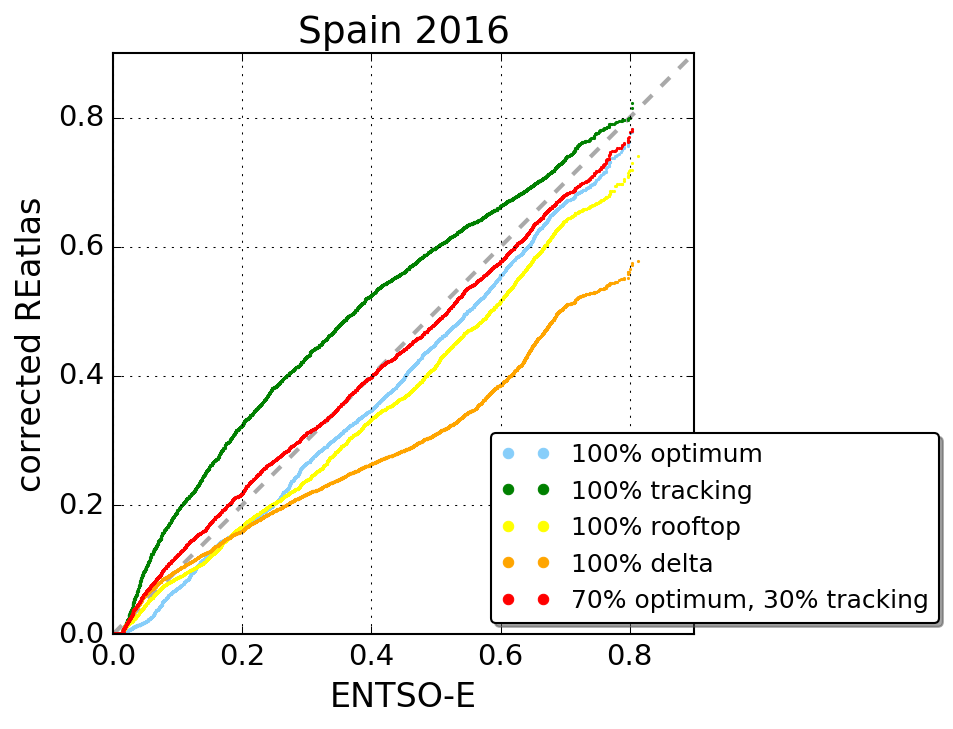}\\
\caption{Duration curve (sorted hourly capacity factor) modeled for different PV system configurations vs. historical data provided by REE \cite{REE_web}. Data for Spain in 2016.} \label{fig_30_70_Spain}
\end{figure}

The dataset comprising the time series for the four configurations is under open license and can be downloaded from the Zenodo repository. Then, to obtain the PV hourly capacity factors for a country, different assumptions on the share of the alternative configurations can be made and the weighted time series can be aggregated accordingly. For instance, Figure \ref{fig_30_70_Spain} shows the duration curves for Spain in 2016 assuming different configurations. They are plotted against duration curve obtained using historical data \cite{REE_web}. A very good agreement is found when the duration curves for optimum tilt and tracking are added assuming 70\%/30\% proportion which roughly represent the current configuration of PV plants in Spain.

\

\begin{figure}[ht!]
\centering
\includegraphics[width=\columnwidth]{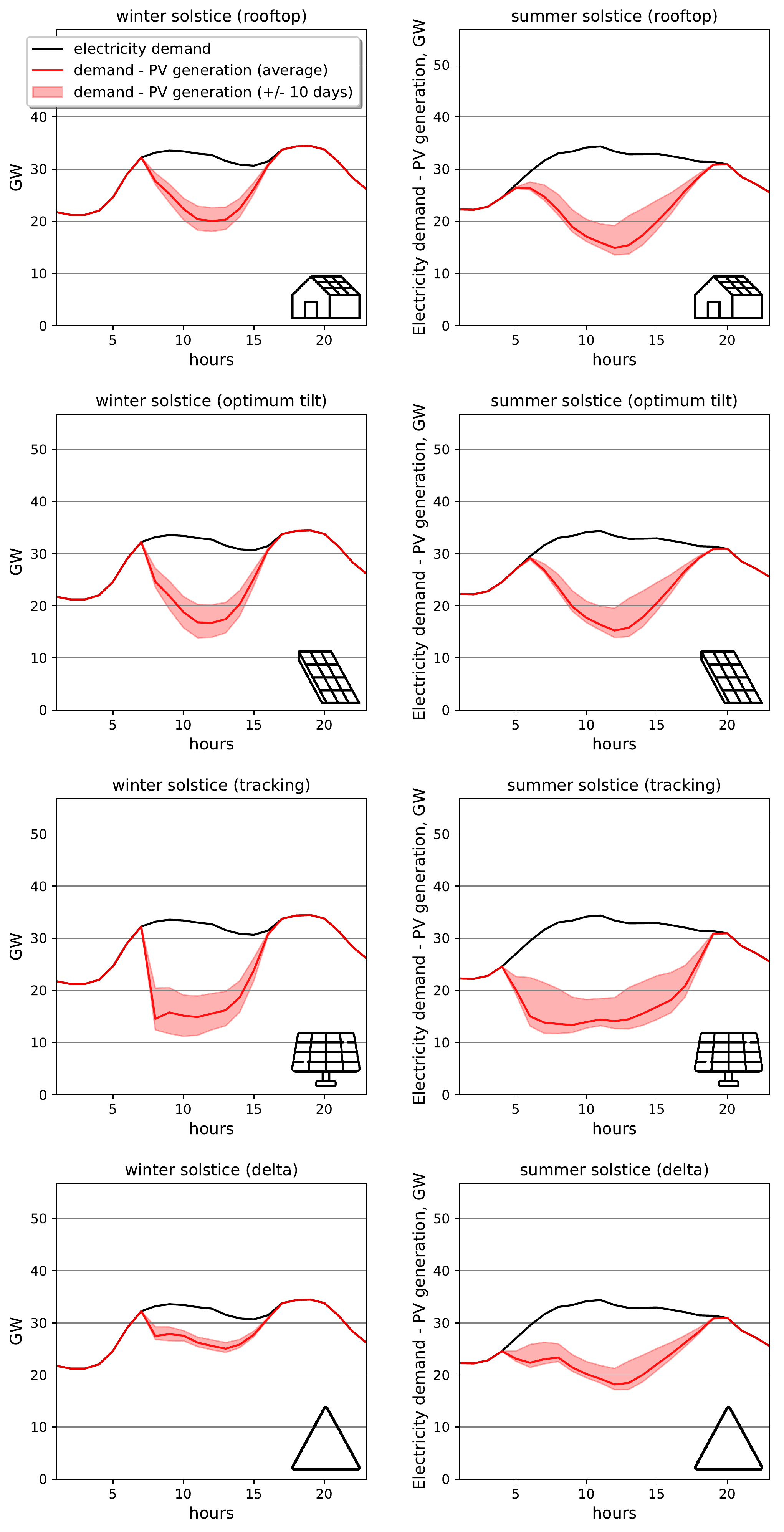}\\
\caption{Electricity demand minus PV generation for winter and summer solstice in Spain, assuming different configurations for the PV panels in the country. The installed PV capacity is assumed to be equal to the yearly-average hourly load (28.4 GW)} \label{fig_duck_curves}
\end{figure}

The produced time series have also been used to assess the impact of PV on the hourly operation of electricity system when large capacities are deployed. The mismatch curve, \textit{i.e.}, electricity demand minus PV generation, is usually employed to investigate this effect. The mistmatch curve is also known in the literature as the `duck curve' \cite{Denholm_2015}. 
%Pfenninger and Staffel \cite{Pfenninger_2016} analyzed the effect that various ammounts of PV installed capacities have on the 'duck-curves (assuming configuration a) and showed that, for some countries, the impact of PV on demand curve is not very significant in winter while mismatch curve can reach negative values in summer. 
Figure \ref{fig_duck_curves} depicts the evolution of the mismatch curve in Spain throughout the winter and summer solstices when a PV capacity equals to the yearly-averaged hourly load (av.h.l.) is installed. In the case of Spain, this corresponds to 28.4 GW. The red line represents the average value for the $\pm$10 days around the solstice in 2015, while the red area includes the minimum and maximum values obtained within that time period. Equivalent curves for other countries in Europe are provided in the Supplementary Materials. The mismatch curves are shown for the four PV configurations previously described. The operation of power systems with high PV penetration requires that throughout the evening, backup technologies increase their production to counterbalance the drop in PV generation. Several strategies have been proposed to re-shape the curve and allow more PV on the grid such us changing operational practices in power system to enable more frequent plant cycling or demand shifting \cite{Denholm_2015}. As Figure \ref{fig_duck_curves} shown, the configuration of PV directly impact the ramp rate and range of backup generation necessary throughout the evening and, consequently, selecting a proper mix of different configurations could ease the operation of the power system when large PV capacities are achieved.

\section{Conclusions}
\label{conclusions}

The research questions that we raised in the introduction can now be answered affirmatively. Irradiance from CFSR reanalysis can be used to obtain PV time series integrated over large-scale regions with a reasonable accuracy. To that end, the bias in reanalysis irradiance must be corrected. We propose a methodology that doesn't require using historical PV output data and, consequently, can be universally applied. When compared to historical PV hourly capacity factors at national scale an error lower or similar to Renewables Ninja and EMHIRES time series, both relaying on satellite-based SARAH irradiance as input, was achieved. 

\

Using global reanalysis irradiance to model PV generation at country level enables the production of consistent long-term time series dataset, including other renewables such as wind and hydroelectricity that can be used as inputs for energy models. The validated methodology was used to produce open-license long-term country-wise PV time series for European countries under four different assumptions for the PV systems configurations (rooftop, optimum tilt, tracking, and delta).

\section{Acknowledgments}
The authors are fully or partially funded by the RE-INVEST project, which is supported by  the  Innovation  Fund  Denmark  under  grant  number  6154-00022B. The responsibility for the contents lies solely with the authors.

\section{References}
\bibliography{energy}

\appendix

\section{Solar conversion in REatlas} \label{annex_solar_conversion}

This annex provides details on the implemented methodology to obtain hourly capacity factors for PV  generation at a national scale. The global renewable energy atlas (REatlas) from Aarhus University \cite{Andresen_2015} uses as input the Climate Forecast System Reanalysis (CFSR) dataset from the National Center for Environmental Prediction (NCEP) \cite{CFSR}. CFSR dataset comprises a 38 years-long global high-resolution dataset (hourly time resolution and approximately 40 x 40 km$^2$ space resolution). The procedure to obtain PV hourly capacity factors representative for a country comprises the following steps.

\begin{enumerate}
\item	At every point in the CFSR grid, the time series for temperature $T_{amb}(t)$ and global horizontal irradiance $G(0,t)$ at ground level\footnote{In the CFSR terminology, $G(0,t)$ is named \textit{downward shortwave radiation at surface}.} are used to compute the irradiance at the entrance of the PV panel and its electricity output. The former is calculated as described in Subsection A.1 (either fixed tilted surface or tracking can be assumed) and the latter is computed using the model for the PV panel described in Section A.2.

\item	For every country $s$ and year $y$, a capacity layout $\left\{C_{n}\right\}_{y,s}$ is built representing the cumulative PV capacity installed at every point $n$ of the CFSR grid. In addition, the configuration of the PV panels in the country under study is also needed. For instance, it can be described by means of two normal distributions, \textit{i.e.}, the panels tilt angles $\beta$ are assumed to follow a Gaussian distribution with mean $\mu_{\beta}$ and standard deviation $\sigma_{\beta}$ while the panels orientation angles $\alpha$ are assumed to follow a Gaussian distribution with mean $\mu_{\alpha}$ and standard deviation $\sigma_{\alpha}$. In the northern hemisphere, $\mu_{\alpha}$ can be assumed to be zero, that is, in average panels are south oriented.

\end{enumerate}

Finally, the capacity factors time series for all the panel configurations and all the grid points $CF_{n}(t, \beta, \alpha)$ are aggregated together to obtain a time series representative for the PV electricity generation in a certain country.

\begin{equation}
CF_{y,s}(t)= \sum_{n, \beta, \alpha} \left\{ C_{n} \right\}_{y,s} CF_{n}(t, \beta, \alpha) PDF(\mu_{\beta}, \sigma_{\beta}) PDF(\mu_{\alpha}, \sigma_{\alpha})
\end{equation}
where $C_{n} \in \left\{C_{n}\right\}_{y,s}$

In this paper, a uniform capacity layout has been assumed for every country, \textit{i.e.}, one solar panel is  installed in every CFSR grid point. 

\subsection{Irradiance on a tilted surface} \label{subsec_irradiance_tilted}
This Section includes a description of the approach followed to: (a) determine the clearness index of the atmosphere for every hour; (b) decompose the global horizontal irradiance at ground level into direct and diffuse irradiance; and (c) estimate the direct, diffuse, and albedo irradiances on a tilted surface and their aggregate value. 

\begin{figure}[ht!]
%\centering
\includegraphics[width=\columnwidth]{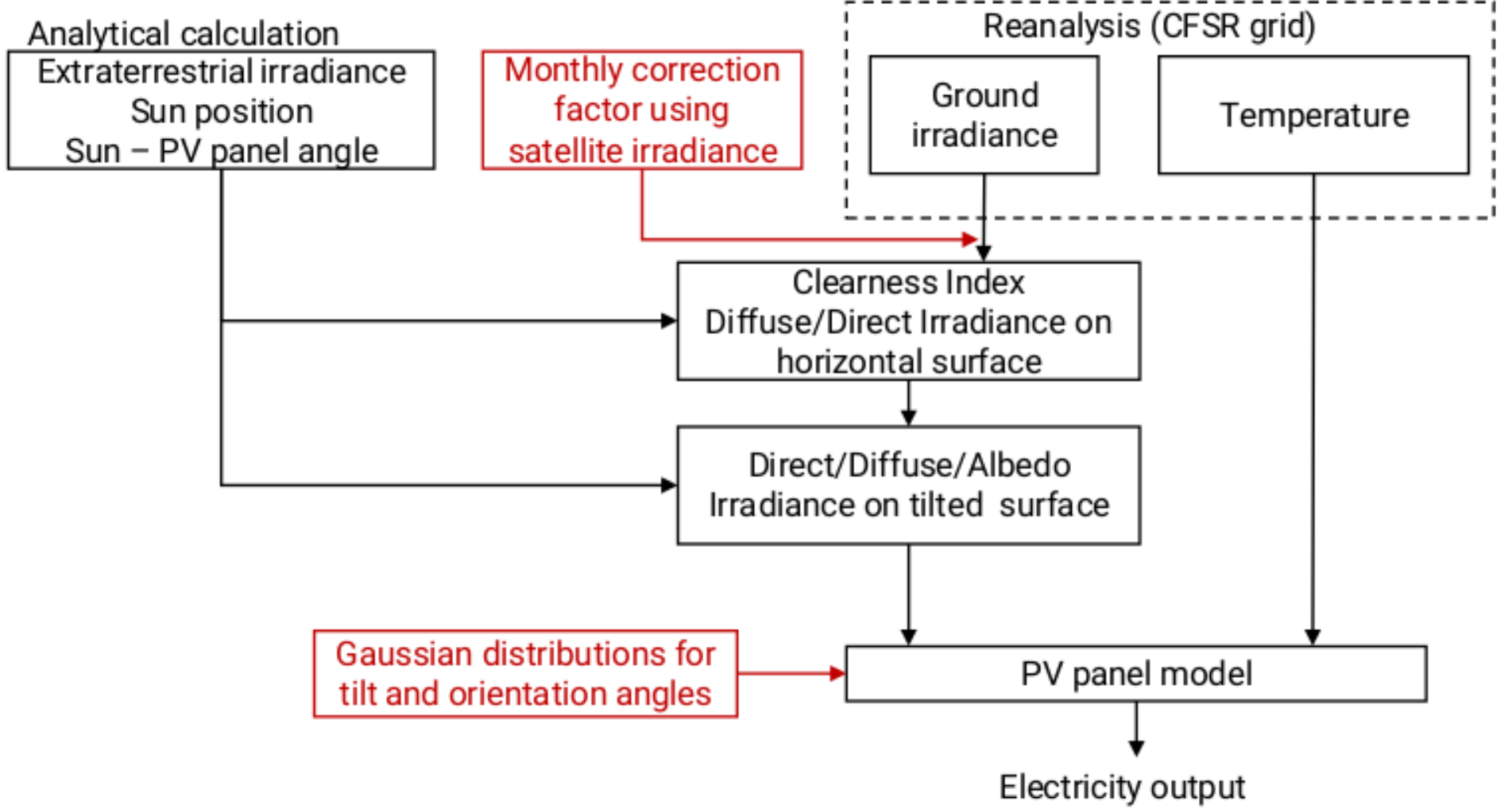}\\
\caption{Scheme of the methodology followed to calculate the electricity output of a PV panel based on the temperature
and irradiance data from reanalysis.} \label{fig_scheme}
\end{figure}

\subsubsection*{Clearness Index}
The extraterrestrial irradiance $B_0(t)$, \textit{i.e.}, the irradiance at the top of the atmosphere, travels through the atmosphere where it is absorbed, reflected and scattered. The clearness index $K_t$ measures the effect of the atmosphere and is influenced by time, date, and location, as well as by the atmosphere composition and cloud content.
\begin{equation} \label{eq_clearness_index}
	K_t=\frac{G(0,t)}{B_0(t)}
\end{equation}

The CFSR provides hourly values for G(0,t) at ground level. $B_0(t)$ can be analytically calculated using the following equation.

\begin{equation}
B_0(t)=B_0 \epsilon(t) \sin \gamma_s
\end{equation}

where $B_0$ is the solar constant $B_0$=1367 W/m$^2$. The eccentricity $\epsilon$ is calculated using equation \ref{eq_eccentricity}, and $\gamma_s$ is the solar altitude obtained from equation \ref{eq_solar_altitude}. 

\

For a certain location on the surface of the Earth with latitude $\phi$ and longitude $\xi$, the position of the Sun in the sky, at any moment, can be described by the solar altitude $\gamma_s$ (angle between the radio-vector of the Sun and the horizontal plane) and the solar azimuth $\psi_s$ (angle between the projection of the Sun radio-vector over the horizontal plane and the south direction). $\gamma_s$ and $\psi_s$ depend on the time and date trough equations \ref{eq_declination} and \ref{eq_ST} .

\begin{equation} \label{eq_solar_altitude}
\sin \gamma_s = \sin \delta \sin \phi + \cos \delta \cos \phi \cos \omega
\end{equation}

\begin{equation}
\cos \psi_s = \frac{(\sin \gamma_s \sin \phi - \sin \delta)}{\cos \gamma_s \cos \phi}[sign(\phi)]
\end{equation}

The declination $\delta$ is computed using the number of the day $d_n$ (counted from the first day of the year) 
\begin{equation} \label{eq_declination}
\delta = 23.45^{\circ} \sin(\frac{360(d_n+284)}{365})
\end{equation}

The true solar time $ST$, expressed in hours, can be calculated with the formula

\begin{equation}
	ST = LT + \frac{ET}{60} - \frac{\xi}{15}
\end{equation}

where $LT$ is the hour of the day (expressed using Universal Coordinated Time, UCT), $\xi$ is the longitude and $ET$ is a correction for the different lengths of the days in a year due to the fact the Earth orbit is not circular but elliptical.

\begin{equation}
	ET = 9.87 \sin (2B) -7.53 \cos(B) -1.5 \sin (B)
\end{equation}
where
\begin{equation}
	B = (d_n -81) \frac{360}{364}
\end{equation}

The true solar time $\omega$ can also be expressed as an angle. In this case $\omega=0$ corresponds to the moment when the Sun is at the highest position, that is, $ST=12$.

\begin{equation} \label{eq_ST}
	\omega = 15^{\circ}(ST -12)
\end{equation}

Finally, the eccentricity $\epsilon$ is calculated as
\begin{equation} \label{eq_eccentricity}
\epsilon = 1 + 0.033 \cos (\frac{360 d_n}{365}) 
\end{equation}

\subsubsection*{Decomposition of global irradiance into direct and diffuse irradiance}

The global horizontal irradiance on a horizontal surface G(0), either measured or, as in this case, modeled, includes the direct horizontal irradiance B(0), that is, the irradiance that reaches the horizontal surface from the Sun without being scattered and the diffuse horizontal irradiance D(0), that is, the irradiance that reaches the horizontal surface after being scattered in the atmosphere. 
The diffuse fraction of the global irradiance $F$ = D(0)/G(0) is estimated based on the clearness index $K_t$ as initially proposed by Liu and Jordan \cite{Liu_1960}. The piecewise function proposed by Reindlt \cite{Reindl_1990}, which is similar to the classic Page model \cite{Page_1961}, is used. 

\begin{equation} \label{eq_F}
	F=
	\
    \begin{cases}
		\min (1, 1.02-0.254 K_t + 0.0123 \sin \gamma_s) , \\
   	\hspace{4cm} \text{if}\ 0 \leq K_t \leq 0.3 \\
		\min (0.97, \max(0.1, 1.4 - 1.749 K_t + 0.177 \sin \gamma_s) , \\
		\hspace{4cm} \text{if}\ 0.3 < K_t \leq 0.78 \\
    \max (0.1, 0.486  K_t - 0.182 \sin \gamma_s) , \\
		\hspace{4cm} \text{if}\ K_t > 0.78 \\
    \end{cases}
\end{equation}

\subsubsection*{Irradiance on a tilted surface}

The irradiance $G(\beta, \alpha)$ on a surface with tilt angle $\beta$ and orientation $\alpha$ is calculated by summing the direct $B(\beta,\alpha)$, diffuse $D(\beta, \alpha)$ and albedo irradiance $R(\beta, \alpha)$.

\begin{equation} \label{eq_G_B_D}
	G(\beta, \alpha) = B(\beta, \alpha) + D(\beta, \alpha) + R(\beta, \alpha)
\end{equation}

If B(0) is known, the direct irradiance B($\beta$, $\alpha$) can be calculated by straightforward geometrical considerations 
\begin{equation}
	B(\beta, \alpha)=\frac{B(0) \max(0, \cos \theta_s)}{\sin \gamma_s}
\end{equation}

where $\gamma_s$ is the solar altitude (see equation \ref{eq_solar_altitude}) and $\theta_s$ is the angle that forms the radio-vector of the Sun and the normal of the surface.

\begin{equation}
\begin{split}
	\cos \theta_s = \sin \delta \sin \phi \cos \beta \\
	& \hspace{-2cm} - [sign(\phi)] \sin \delta \cos \phi \sin \beta \cos \alpha \\
	& \hspace{-2cm}  + \cos \delta \cos \phi \cos \beta \cos \omega \\
	& \hspace{-2cm} + [sign(\phi)] \cos \delta \sin \phi \sin \beta \cos \alpha \cos \omega \\
	& \hspace{-2cm} + \cos \delta \sin \alpha \sin \omega \sin \beta
	\end{split}
\end{equation}

The diffuse irradiance $B(\beta, \alpha)$ is calculated using the anisotropic model proposed by Hay and Davies \cite{Hay_1985}, where $D(\beta, \alpha)$ is composed of a circumsolar component coming directly from the direction of the Sun $D_{circumsolar}(\beta,\alpha)$, and an isotropic component coming from the entire celestial hemisphere $D_{isotropic}(\beta,\alpha)$. The anisotropy index $k_1$ weights both components. $k_1$ is estimated by the ratio of the direct irradiance on the ground $B(0)$ and at the top of the atmosphere $B_0(0)$.

\begin{equation}
	k_1=\frac{B(0)}{B_0(0)}
\end{equation}

\begin{equation} \label{eq_iso_circum}
	D(\beta,\alpha)=D_{circumsolar}(\beta,\alpha)+D_{isotropic}(\beta,\alpha) 
\end{equation}

where

\begin{equation}
	D_{circumsolar}(\beta,\alpha)= k_1 \frac{D(0) \max(0, \cos \theta_s)}{\sin \gamma_s}
\end{equation}

and the classic definition of isotropic diffuse irradiance $D_{isotropic}(\beta,\alpha)$ has been modified to include
the horizon brightening effect $k_{horizon}$ as proposed in \cite{Reindl_1990}.

\begin{equation}
	D_{isotropic}(\beta,\alpha)= k_{horizon} (1-k_1) D(0) \frac{1 + \cos \beta}{2}
\end{equation}

\begin{equation}
	k_{horizon}=1+\sqrt{k_1} \sin^3(\frac{\gamma_s}{2})
\end{equation}

Finally, the albedo irradiance $R(\beta,\alpha)$ is calculated as

\begin{equation}
	R(\beta,\alpha)= \rho G(0) \frac{1 - \cos \beta}{2}
\end{equation}

where $\rho$ is the reflectivity of the ground. $\rho$ is determine for every grid point using the information provided by the CFSR database which includes downward shortwave radiation at surface, \textit{i.e.} G(0), and upward shortwave at surface R(0).

\begin{equation}
	\rho= \frac{R(0)}{G(0)}
\end{equation}

\subsection{Model of the PV panel} \label{subsec_pv_panel}

The REatlas includes a simple model for the PV panel whose parameters can be obtained from commercial data sheet of PV panels. 
It is worth noticing that the conversion efficiency $\eta_{STC}$ assumed for the PV panel under Standard Test Conditions (STC) does not have any influence on the hourly capacity factor time series. In fact, $\eta_{STC}$ is only necessary if one wants the compute the area that must be covered with PV panels in order to install a certain capacity. Conversely to wind capacity factors, where the turbine power curve directly influences the time series, PV capacity factors are mainly influenced by the configuration of the panels (tilt and orientation angles if fixed or tracking strategy) but not by $\eta_{STC}$.

The temperature dependence of the efficiency has been included in the model using the efficiency thermal coefficient $\gamma_T$ .The efficiency is then computed as
\begin{equation} \label{eq_TONC}
	\eta = \eta_{STC} [1-\gamma_T(T_{cell}- T_{cell, STC})]
\end{equation}
where $T_{cell, STC}$ is the temperature of the cell under Standard Test Conditions. For silicon flat panels, $\gamma_T$ = -0.4\%/$^{\circ}C$ is assumed.

\

$T_{cell}$ is calculated using the Nominal Operating Cell Temperature (NOCT). NOCT conditions include $G_{NOCT}$= 800 W/m$^2$ and $T_{amb, NOCT}$=20$^{\circ}C$. From the ambient temperature at every grid point $T_{amb}$,  $T_{cell}$ can be computed applying the energy balance equation to the PV module.

\begin{equation}
\begin{split}
	T_{cell} = T_{amb} \\
	& \hspace{-2cm}+ \frac{G}{G_{NOCT}} (T_{cell, NOCT}- T_{amb, NOCT}) (1-\frac{\eta}{0.9})
\end{split}
\end{equation}

Finally, the model includes a system efficiency $\eta_{system}$=0.9 which represents the aggregate effect of wiring losses, inverter efficiency, and voltage conversion efficiency. 0.9 is the average value found only for the inverter efficiency for all the individual sites processed in \cite{Pfenninger_2016}. However, the inverter efficiency of new installation is expected to be better \cite{report_Fraunhofer}. Hence, 0.9 is selected as a coefficient representing the aggregate efficiency of the system (wiring, inverter and voltage conversion). The possible curtailment due to PV generated power exceeding the rated power of the inverted is not included in the model. The hourly capacity factors can be then computed as the ratio of the energy produced at any hour and the generation under Standard Test Conditions (STC).

\begin{equation}
\begin{split}
CF_n(t, \beta, \alpha) = \eta_{system} \frac{G_n(t,\beta,\alpha)\eta(T_{amb})}{G_{STC}\eta (T_{STC})} \\
= 0.9 \frac{G_n(t,\beta,\alpha)\eta(T_{amb})}{1000 W/m^2 \eta (298K)}
\end{split}
\end{equation}

\section{Monthly correction factors} \label{annex_monthly_correction_factors}

\begin{table*}[!htbp]
  \centering
  \caption{Monthly correction factors $C_{m,s}$ for European countries}
	\begin{footnotesize}
   
    \begin{tabular}{lrrrrrrrrrrrr}
		\hline
      & Jan     & Feb     & Mar     & Apr     & May     & Jun     & Jul     & Ago     & Sep     & Oct    & Nov    & Dic \\
		\hline
		
    AUT   & 0.78  & 0.82  & 0.84  & 0.87  & 0.86  & 0.85  & 0.88  & 0.91  & 0.89  & 0.89  & 0.81  & 0.79 \\
    BEL   & 1.04  & 0.90  & 0.94  & 0.94  & 0.91  & 0.93  & 0.94  & 0.94  & 0.98  & 1.02  & 1.07  & 1.09 \\
    BGR   & 0.85  & 0.78  & 0.87  & 0.89  & 0.90  & 0.94  & 0.96  & 0.99  & 0.96  & 0.95  & 0.89  & 0.88 \\
		BIH	  & 0.88	& 0.83	& 0.85	& 0.86	& 0.84	& 0.89	& 0.91	& 0.94	& 0.91	& 0.92	& 0.90	& 0.87 \\
    CHE   & 0.62  & 0.67  & 0.78  & 0.86  & 0.84  & 0.87  & 0.88  & 0.90  & 0.89  & 0.81  & 0.73  & 0.63 \\
    CYP   & 0.95  & 0.92  & 0.94  & 0.94  & 0.96  & 0.98  & 0.98  & 0.98  & 0.98  & 0.96  & 0.97  & 0.97 \\
    CZE   & 0.84  & 0.80  & 0.84  & 0.89  & 0.86  & 0.88  & 0.88  & 0.89  & 0.93  & 0.92  & 0.87  & 0.90 \\
    DEU   & 0.97  & 0.87  & 0.91  & 0.90  & 0.88  & 0.90  & 0.90  & 0.91  & 0.95  & 0.97  & 1.00  & 1.00 \\
    DNK   & 0.90  & 0.84  & 0.91  & 0.93  & 0.91  & 0.91  & 0.93  & 0.92  & 1.00  & 1.01  & 1.08  & 0.95 \\
    ESP   & 1.00  & 0.99  & 0.97  & 0.97  & 0.96  & 0.97  & 0.98  & 0.98  & 0.98  & 0.98  & 1.02  & 1.00 \\
    EST   & 0.79  & 0.84  & 1.01  & 0.95  & 0.92  & 0.91  & 0.90  & 0.89  & 0.97  & 1.07  & 0.97  & 0.85 \\
    FIN   & 0.80  & 0.84  & 1.02  & 1.12  & 1.00  & 0.94  & 0.89  & 0.91  & 0.98  & 1.01  & 0.91  & 0.81 \\
    FRA   & 0.96  & 0.93  & 0.94  & 0.94  & 0.91  & 0.93  & 0.95  & 0.96  & 0.97  & 0.96  & 1.01  & 1.01 \\
    GBR   & 1.02  & 0.96  & 0.96  & 0.96  & 0.95  & 0.99  & 0.98  & 1.07  & 1.08  & 1.06  & 1.11  & 1.03 \\
    GRC   & 0.93  & 0.87  & 0.90  & 0.91  & 0.93  & 0.98  & 1.01  & 1.00  & 0.97  & 0.94  & 0.93  & 0.92 \\
    HRV   & 0.92  & 0.86  & 0.91  & 0.90  & 0.90  & 0.93  & 0.96  & 0.96  & 0.94  & 0.94  & 0.91  & 0.92 \\
    HUN   & 0.88  & 0.79  & 0.92  & 0.92  & 0.91  & 0.95  & 0.97  & 0.96  & 0.96  & 0.98  & 0.94  & 0.89 \\
    IRL   & 1.06  & 1.02  & 1.03  & 0.98  & 1.03  & 1.03  & 1.10  & 1.30  & 1.22  & 1.12  & 1.10  & 1.10 \\
    ITA   & 0.91  & 0.87  & 0.90  & 0.90  & 0.91  & 0.94  & 0.97  & 0.98  & 0.95  & 0.93  & 0.91  & 0.91 \\
    LTU   & 0.94  & 0.90  & 0.92  & 0.95  & 0.92  & 0.94  & 0.93  & 0.94  & 0.99  & 1.09  & 1.08  & 1.02 \\
    LUX   & 1.02  & 0.93  & 0.97  & 0.96  & 0.93  & 0.93  & 0.93  & 0.95  & 0.97  & 0.99  & 1.05  & 1.15 \\
    LVA   & 0.89  & 0.92  & 0.95  & 0.98  & 0.92  & 0.93  & 0.93  & 0.94  & 0.99  & 1.10  & 1.04  & 0.96 \\
    MLT   & 0.95  & 0.93  & 0.94  & 0.96  & 0.98  & 0.99  & 1.00  & 1.00  & 0.99  & 0.96  & 0.94  & 0.95 \\
    NLD   & 1.09  & 0.95  & 0.92  & 0.93  & 0.90  & 0.95  & 0.94  & 0.95  & 1.02  & 1.05  & 1.14  & 1.11 \\
    NOR   & 0.57  & 0.73  & 0.87  & 0.98  & 0.97  & 0.97  & 0.94  & 0.93  & 0.97  & 0.92  & 0.77  & 0.56 \\
    POL   & 0.96  & 0.87  & 0.89  & 0.93  & 0.91  & 0.93  & 0.93  & 0.92  & 0.98  & 1.00  & 1.03  & 1.04 \\
    PRT   & 1.05  & 1.01  & 0.97  & 0.96  & 0.96  & 0.96  & 0.97  & 0.97  & 0.98  & 0.98  & 1.01  & 1.03 \\
    ROU   & 0.80  & 0.76  & 0.87  & 0.89  & 0.91  & 0.91  & 0.95  & 0.95  & 0.96  & 0.95  & 0.89  & 0.80 \\
		SRB	  & 0.83	& 0.74	& 0.86	& 0.87	& 0.88	& 0.91	& 0.94	& 0.95	& 0.92	& 0.93	& 0.90  & 0.84 \\
    SVK   & 0.90  & 0.86  & 0.90  & 0.90  & 0.86  & 0.88  & 0.90  & 0.90  & 0.95  & 0.94  & 0.94  & 0.92 \\
    SVN   & 0.92  & 0.89  & 0.93  & 0.89  & 0.90  & 0.92  & 0.94  & 0.94  & 0.93  & 0.95  & 0.96  & 0.89 \\
    SWE   & 0.80  & 0.81  & 0.92  & 1.01  & 0.96  & 0.93  & 0.91  & 0.90  & 0.95  & 0.96  & 0.91  & 0.82 \\

		\hline
		\end{tabular}
		\end{footnotesize}
  \label{table_monthly_factor}
\end{table*}

\FloatBarrier

\section{Gaussian distributions for tilt and orientation angles} \label{annex_gaussian_parameters} 

\begin{table}[!htbp]
  \centering
  \caption{Set of parameters to describe the inclination and orientation angles of PV panels in every country.}
	\begin{footnotesize}

    \begin{tabular}{lrrrrr}
		\hline
    country & $\mu_{\beta}$ & $\sigma_{\beta}$ & $\mu_{\alpha}$ & $\sigma_{\alpha}$ & $\zeta$ \\
		\hline
    AUT   & 40    & 20    & 0     & 30    & 1 \\
    BEL   & 10    & 20    & 0     & 30    & 0.8 \\
    BGR   & 40    & 20    & 0     & 30    & 0.75 \\
    CZE   & 25    & 20    & 0     & 30    & 0.7 \\
    DEU   & 20    & 20    & 0     & 30    & 0.65 \\
    DNK   & 20    & 20    & 0     & 30    & 0.95 \\
    ESP   & 10    & 20    & 0     & 30    & 1 \\
    FRA   & 20    & 20    & 0     & 30    & 0.8 \\
    GRC   & 40    & 20    & 0     & 30    & 0.85 \\
    ITA   & 35    & 20    & 0     & 30    & 0.65 \\
    LTU   & 10    & 20    & 0     & 30    & 1 \\
    NLD   & 40    & 20    & 0     & 30    & 0.95 \\
    PRT   & 25    & 20    & 0     & 30    & 1 \\
    ROU   & 35    & 20    & 0     & 30    & 0.7 \\
    SVK   & 10    & 20    & 0     & 30    & 1 \\
    SVN   & 10    & 20    & 0     & 30    & 1 \\
		\hline
    \end{tabular}
   
		\end{footnotesize}
  \label{table_monthly_factor}
\end{table}

\end{document}